\documentclass[12pt,draftcls,onecolumn]{IEEEtran}
\usepackage{cite}
\usepackage{amsmath,amssymb,amsfonts}
\usepackage{algorithm}
\usepackage{algorithmic}

\usepackage{graphicx}
\usepackage{textcomp}
\usepackage{xcolor}
\usepackage{subfigure}
\usepackage{url}
\usepackage{amsmath}

\newtheorem{theorem}{Theorem}
\newtheorem{lemma}{Lemma}

\begin{document}

\title{Deep Learning for Asynchronous Massive Access with Data Frame Length Diversity}


\author{Yanna~Bai,
        Wei~Chen,~\IEEEmembership{Senior~Member,~IEEE,}
        Bo~Ai,~\IEEEmembership{Fellow,~IEEE,}
        Petar~Popovski,~\IEEEmembership{Fellow,~IEEE}
\thanks{Yanna Bai, Wei chen and Bo Ai are with the State Key Laboratory of Rail Traffic Control and Safety, Beijing Jiaotong University, Beijing 100044, China, Beijing 100044, China. (Email: {yannabai, weich, boai}@bjtu.edu.cn).}
\thanks{Petar Popovski is with Aalborg University, Denmark. (Email: petarp@es.aau.dk).}}

\maketitle

\begin{abstract}
Grant-free non-orthogonal multiple access has been regarded as a viable approach to accommodate access for a massive number of machine-type devices with small data packets. The sporadic activation of the devices creates a multiuser setup where it is suitable to use compressed sensing in order to detect the active devices and decode their data. We consider asynchronous access of machine-type devices that send data packets of different frame sizes, leading to \emph{data length diversity}. We address the composite problem of activity detection, channel estimation, and data recovery by posing it as a structured sparse recovery, having three-level sparsity caused by sporadic activity, symbol delay, and data length diversity. We approach the problem through approximate message passing with a backward propagation algorithm (AMP-BP), tailored to exploit the sparsity, and in particular the data length diversity. Moreover, we unfold the proposed AMP-BP into a model-driven network, termed learned AMP-BP (LAMP-BP), which enhances detection performance. The results show that the proposed LAMP-BP outperforms existing methods in activity detection and data recovery accuracy.
\end{abstract}


\IEEEpeerreviewmaketitle

\section{Introduction}

\IEEEPARstart{M}{assive} machine-type communication (mMTC) emerged as a part of 5G and will remain a consituent of B5G/6G towards supporting a vast number of  machine-type devices (MTDs) for Internet-of-things applications \cite{8766143, 9464917}. Besides large number of devices, mMTC features low data rates, small data packets, and sporadic activity, which requires tailored access schemes. The existing approaches to mMTC access can be, roughly, divided into two classes: evolutionary grant-based access, which builds upon 4G networks, and the revolutionary grant-free (GF) access~\cite{9320554}. Compared to grant-based access, GF-based random access (GF-RA) omits the complex handshaking process and devices could directly transmit packets without waiting for signaling exchange \cite{9537931}. Thus GF-RA has a low signaling overhead and low access latency. The combination of GF-RA and non-orthogonal multiple access (NOMA) further enhances the network access capability \cite{9097306,9031550,8482464,9140386}. The joint consideration of user activity detection, channel estimation and inter-user interference makes GF-NOMA more challenging, and thus calls for advanced system design and intelligent signal processing.

\par
A potential solution would be to utilize the sparse nature of mMTC to alleviate the uncertainty due to the unknown user activity \cite{8454392}. In mMTC, the activity of the MTDs is usually sporadic and, in many cases, event-driven. At the same time, only a small portion of devices are active. Thus the activity indicator of all devices is a sparse signal. In compressed sensing (CS) based contention-free GF-RA, each device is assigned a unique pilot. The base station (BS) performs jointly user activity detection and channel estimation with the received pilot signal and known pilot matrix of all devices by solving a sparse recovery problem. This sparse recovery problem can be solved via either traditional model-driven CS algorithms \cite{1614066,4385788} or data-driven CS methods \cite{9903376,9605579,9685696,9252937}.
Model-driven methods solve the optimization problem through a number of iterations. Generally, the convergence speed of model-driven methods is limited, but it does have theoretical guarantees. Data-driven methods use neural networks to learn the mapping between the received signal and the reconstructed signal via a training data set where faster convergence and higher estimation accuracy have been observed empirically~\cite{9903376,7934066}. In CS-based GF-RA, a general way to enhance the detection performance of CS algorithms is by exploring various structures embedded in the estimated signal. For example, when the BS has multiple antennas, the detection problem becomes a multiple measurement vector (MMV) problem in CS \cite{8323218}. In \cite{8323218}, Liu and Yu demonstrated the performance gain from multiple antennas and asymptotically analyze the detection error becomes zero when the number of antennas tends to infinity. In \cite{9566698}, Xiao et al. explored the angular domain sparsity in massive multiple input multiple output (MIMO), where the corresponding MMV problem is the reconstruction problem of a row-sparse matrix with intra-row sparsity. In \cite{9839006}, Bai et al. proposed to use dictionary learning to learn the potential sparsity in other domains and improve the detection performance by reconstructing a sparser signal. MMV problems also exist even when there is a single antenna. In \cite{9364871}, Jiang and Wang considered the temporal correlation in the adjacent time step and constructed the activity detection and channel estimation as an MMV problem. In \cite{9268113}, Xiao et al. considered joint activity detection, channel estimation, and data recovery into an MMV problem, while further considering the special structure induced by the data length diversity.

\par
However, most of those works assume that the transmission of active devices are perfectly synchronized. This assumption is unreasonable in GF-RA, as active devices directly transmit their data without timing advance. Several works \cite{9413870,8716690,9814672,9390399} consider the massive random access case where synchronization is not completely lost, i.e., the frames of different users are asynchronous, but the symbols are still synchronized. In \cite{9413870,8716690,9814672,9390399}, the unsynchronized sporadic transmission in mMTC is formulated into an estimation problem with a special block sparse structure. Traditional model-driven CS algorithms \cite{9413870,8716690,9814672} and data-driven methods \cite{9390399} were used to explore this block-sparse structure towards user activity detection and channel estimation. Plenty of research has shown the potential of CS-based GF-RA for mMTC. Yet, the previous contention-free CS-based access are limited in terms of scalability, as a unique preamble or pilot sequence needs to be assigned to each device, The number of devices in simulations is, generally, limited to a number lower than the targeted thousands of nodes, while the dimension of the activity detection problem increases dramatically with the increase of the number of devices. Contention-based GF-RA \cite{9267798} is more suitable for massive access, as it decreases the computation complexity at the receiver by adopting the shared pilots for all users. Active devices randomly select pilots from the common pool of pilots. Then the receiver performs the pilot detection and channel estimation for data recovery, and identifies active users via the unique ID contained in transmitted data packets. The chance of pilot collision is small when if the number of active users is below some threshold. To the best of our knowledge, no work has yet considered contention-based CS-based asynchronous access. One challenge in contention-based access is the ambiguity between pilot detection to active user detection, as different users with different symbol delays may select the same pilot. Existing algorithms \cite{9413870,8716690,9814672,9390399} fail to resolve the collisions in contention-based random access, which degrades the performance.

\par
In this paper, we consider contention-based asynchronous random access with data length diversity. Instead of employing user-specific pilot sequences, in contention-based random access, active devices randomly select sequences from the common pool for access. Thus the access scheme is more flexible to support more devices. Different active users are synchronized at the symbol level, while being asynchronous at the frame level. There is a guard time after the pilot transmission to prevent the interference between pilot and data. The data length diversity comes from the different amount of transmitted data of diverse devices. To the best of our knowledge, this is the first work that considers the contention-based asynchronous random access with data length diversity. We construct the pilot detection, channel estimation, data recovery and activity detection at the receiver as an MMV problem with structural sparsity. Then we propose an approximate message passing with backward propagation (AMP-BP) to solve this structural MMV problem by exploring the data length diversity. Furthermore, we extend the proposed AMP-BP into a data-driven neural network, relying on deep unfolding. Both theoretical analysis and experimental results demonstrate the performance gain of the proposed receiver.

\par
The main contributions are summarized as follows:
\begin{itemize}
  \item We address the problem of joint pilot detection, channel estimation, data recovery, and activity detection in asynchronous contention-based random access with data length diversity by posing it as an MMV problem with three sources of sparsity: user activity, different symbol delay of users, and different number of data symbols.
  \item We propose AMP-BP to solve the structural MMV problem by exploring the data length diversity. Furthermore, we unfold the AMP-BP into a neural network with learn-able parameters, and train the learned AMP-BP (LAMP-BP) in an end-to-end manner. Compared with the original LAMP network, the LAMP-BP has well-designed structure and activation function to explore the special sparse structural in the MMV problem and thus enhance the performance of data recovery.
  \item We demonstrate the performance gain of the proposed algorithm through both theoretical analysis and experiments. The analysis shows that the proposed algorithm enables the recovery of more users with the same communication resources as compared to the benchmark.
\end{itemize}

\par
The rest of this paper is organized as follows. Section II shows the system model of contention-based random access with imperfect synchronization and data length diversity for the case of single and multiple antennas, respectively. In Section III, we pose joint pilot detection, channel estimation, data recovery, and activity detection as an MMV problem with three sparsity sources, and propose a general receiver with iterative CS algorithms and the corresponding model-driven neural network. In Section V, we analyze the performance gains of the proposed method theoretically, while the simulation results are presented in Section V. Section VI concludes the paper.

\par
Notation: Bold letters are used for vectors and matrices, normal letters represent scalars. Mathcal letters like $\mathcal{A}$ represent collections. $\mathbb{C}$ and $\mathbb{R}$ represent the complex number field and the real number field, respectively. $\mathbf{0}_{L}$ represents the all-zero vector with length $L$. Table \ref{notation} summarizes parts of necessary notations.

\begin{table}[h]
\renewcommand\arraystretch{0.8} 
\centering  
\setlength{\abovecaptionskip}{0pt}%
\setlength{\belowcaptionskip}{4pt}%
\caption{Technical Notations.}\label{notation}
\footnotesize
\begin{tabular}{c c} 
\hline
Symbol  &Description   \\ \hline\hline
$N$  &Number of users \\
$N_{a}$ &Number of active users\\
$\mathbf{S}$ &The common spreading matrix\\
$\mathbf{s}_{m}$ &The $m$-th spreading sequence\\
$M$  &Number of common spreading sequences \\
$L_s$  &Length of the spreading sequence \\
$\mathbf{p}_{k} $    &Pilot signal of user $k$   \\
$L_{p}$   &Number of transmitted pilot symbols \\
$\mathbf{d}_{k} $    &Data signal of user $k$   \\
$L_{d}$   &Maximum number of transmitted data symbols    \\
$L_{d}^{k}$   &Number of transmitted data symbols of user $k$ \\
$t_{k}$    &Number of delay symbols of user k    \\
$T_{max}$  &Maximum number of delay symbols   \\
$T_{g}$  &Guard time    \\
$h^{k} $    &Channel of user $k$ under single antenna  \\
$l_{k} $    &Path loss of user $k$ \\
$g^{k} $    &The Rayleigh fading of user $k$ \\
$\mathbf{s}^{k} $    &The spreading sequence selected by user $k$\\
$\mathbf{Y}_{p}$ &Received pilot signal   \\
$\mathcal{K}$  &The index set of active users \\
$\mathbf{\hat{s}}_{t_k}^{k}$ &The expanded spreading sequence of user $k$ \\
$\mathbf{\hat{S}}$  &The expanded spreading matrix\\
$\mathbf{U}$  &The pilot matrix\\
$\mathbf{N}_p$  &Noise in received pilot signal\\
$\mathbf{Y}_{d}$ &Received data signal   \\
$\mathbf{Z}$  &The data matrix\\
$\mathbf{N}_d$  &Noise in received data signal\\
$r$    &Number of antennas  \\
$\mathbf{h}^{k}$    &Channel of user $k$ under multiple antenna  \\
$h^{k,r}$    &Channel of user $k$ at $r$-th antenna  \\
$\mathbf{Y}_{p,r} $  &Received pilot signal at $r$-th antenna\\
$\mathbf{Y}_{d,r} $  &Received data signal at $r$-th antenna\\
$\mathbf{Y}$  &$[\mathbf{Y}_{p},\mathbf{Y}_{d}]$, combination of received pilot and data\\
$\mathbf{X}$  &$[\mathbf{U},\mathbf{Z}]$, combination of pilot matrix and data matrix\\
$\tilde{M}$  &$L_s+T_g$, number of rows of MMV problem\\
$\tilde{N}$  &$N(T_g+1)$, number of columns of MMV problem \\
$L$  &$L_p+L_d$, number of columns of measurement\\
\hline
\end{tabular}
\end{table}

\section{System Model}

We consider an asynchronous massive access scenario with a BS with one or multiple antennas and $N$ single-antenna MTDs (users). In each period, $N_{a}$ active users independently transmit data to the BS, where $N_{a}\ll N$ due to sporadic user activity. Active users directly transmit pilots and data to the BS without waiting for scheduling, while the BS performs joint activity detection, channel estimation, and data recovery with the received pilot and data signals. Differently from general CS-based access schemes where each user is assigned with a unique signature for activity identification, such as pilot sequence \cite{8323218} or spreading sequence \cite{9268113}, we adopt a contention-based access scheme~\cite{9267798} for enhancing scalability. For each cell, there is a common spreading sequence pool $\mathbf{S}=[\mathbf{s}_{1},\mathbf{s}_{2},\ldots,\mathbf{s}_{M}] \in \mathbb{R}^{L_s \times M}$, where $L_s$ is the length of spreading sequence and $M$ is the number of spreading sequences. The active users are identified via the unique user ID contained in data symbols.

\begin{figure*}
\centering
\subfigure[The asynchronous access in time domain.]{
\label{fig:nmsenonnoise}
\includegraphics[width=0.5\textwidth]{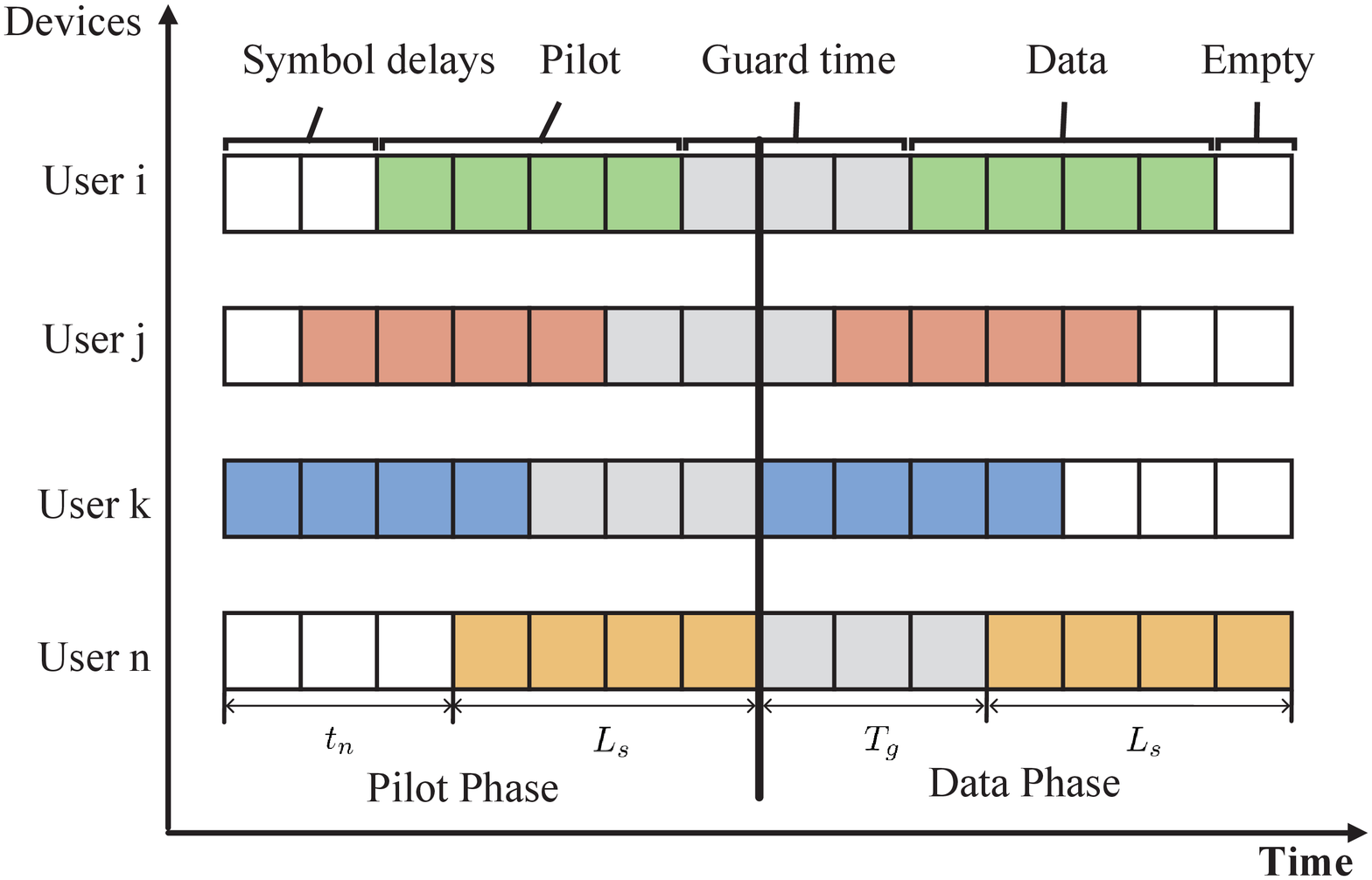}}
\vspace{.1in}
\subfigure[The data length diversity in frequency domain]{
\label{fig:nmsenoise}
\includegraphics[width=0.35\textwidth]{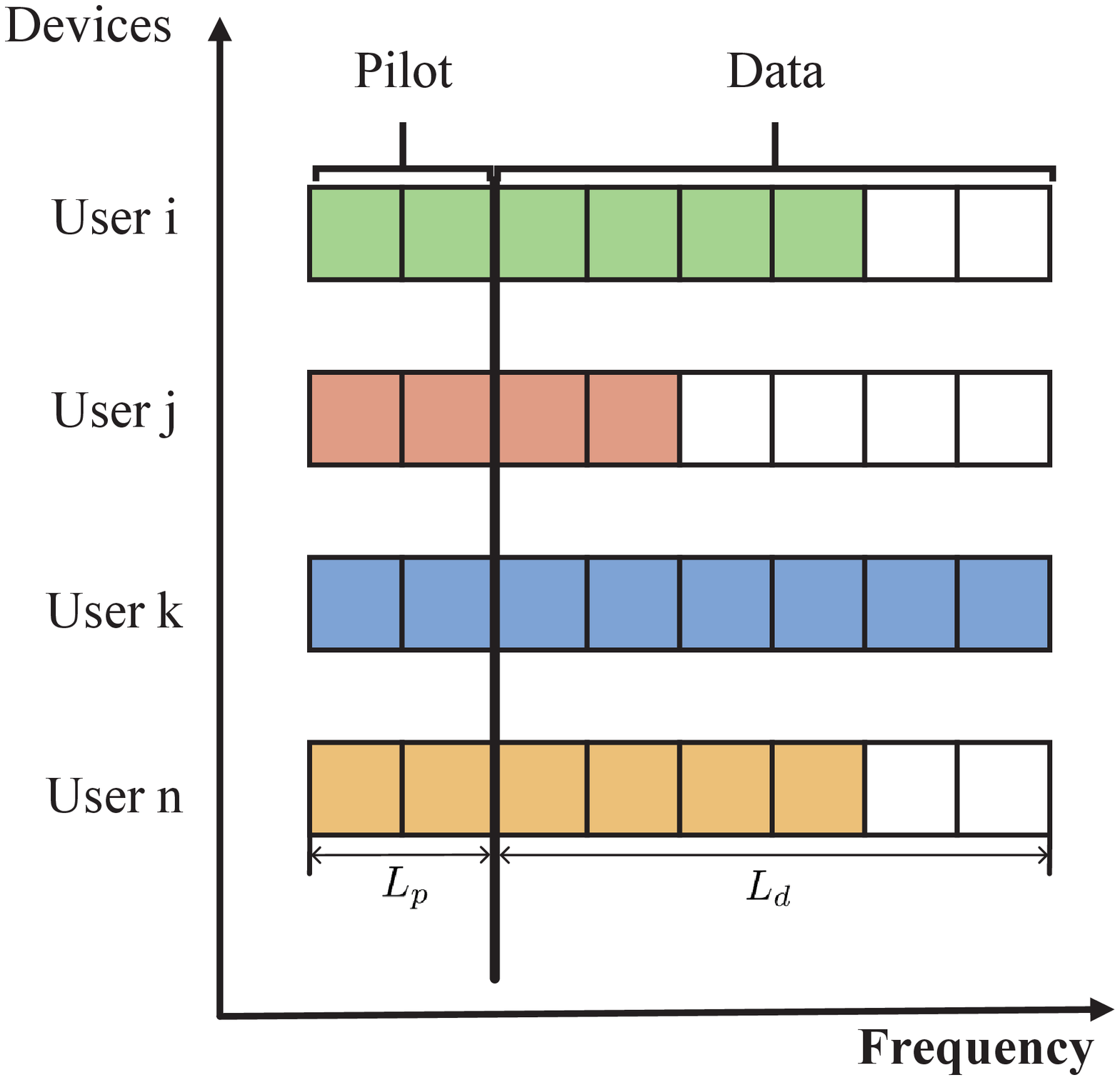}}
\caption{An illustration of the asynchronous massive access with data length diversity. The left figure shows the frame level asynchronicity and the right figure shows the different amount of transmitted data from different devices.}
\label{fig:CS-RA} 
\end{figure*}

We consider the situation where multiple MTDs transmit different amount of data of varying lengths, resulting in data length diversity. The pilot signal of user $k$ is $\mathbf{p}_{k}$, which contains $L_{p}$ repeated pilot symbols for reliable channel estimation. The data signal of user $k$ is $\mathbf{d}_k$, which contains $L_{d}^{k}$ data symbols and $L_{d}-L_{d}^{k}$ zeros indicating no symbol transmission. $L_{d}$ denotes the maximum length of transmitted data symbols. The number of pilot symbols $L_p$ and data symbols $L_d$ is known at the BS, while the exact number of transmitted data symbols of each active user is unknown. This is common in scenarios with diversity of communication services or device types. Data length diversity brings additional complexity, while providing a potential gain for active user detection at the BS. 

We consider asynchronous access, where the symbol delay of active user $k$ at the base station is $t_{k}$, such that $t_{k}$ follows an independent discrete uniform distribution\footnote{We use this simple assumption to introduce maximal uncertainty within a bounded delay.} of $\{0,1,\ldots,T_{max}\}$, and is unknown at the BS. After the pilot transmission, there is a guard interval before the data transmission, see~\cite{9390399,6692494}. The length of guard time is $T_{g}$ and we have $T_{g}\geq T_{max}$ to prevent the overlapping of pilot and data symbols. To simplify, we assume $T_{g}= T_{max}$. Fig.~\ref{fig:CS-RA} shows an illustration of the asynchronous massive access with data length diversity.

\subsection{Single antenna scenario}

\par
For a single-antenna BS, the narrowband channel coefficient of user $k$ is defined as $h^{k}=l_{k}g_{k}$, where $l_{k}$ defines the path loss and $g_{k} \sim \mathcal{CN}(0,1)$. We use $\mathbf{s}^{k}$ to define the spreading sequence selected by user $k$. Then the received pilot signal $\mathbf{Y}_{p}\in \mathbb{C}^{(L_{s}+T_{g})\times L_{p}}$ at the BS is given by
\begin{equation}
\mathbf{Y}_{p} =\sum\limits_{k\in\mathcal{K}}h^{k}\mathbf{\hat{s}}_{t_k}^{k}\mathbf{p}_{k}^{T}+\mathbf{N}_{p}
=\mathbf{\hat{S}}\mathbf{U}+\mathbf{N}_{p},
\label{Y_pilot}
\end{equation}
where $\mathcal{K}\subset\{1,2,\ldots,N\}$ denotes the index of active users, $\mathbf{\hat{s}}_{t_k}^{k}=[\mathbf{0}_{t_k}^{T},(\mathbf{s}^{k})^{T},\mathbf{0}_{T_g-t_k}^{T}]^{T}\in \mathbb{C}^{(L_s+T_g)\times 1} $ denotes the expanded spreading sequence of user $k$ that is obtained by adding $t_k$ and $T_g-t_k$ zeros in the front and end of the spreading sequence $\mathbf{s}^{k}$ to model the symbol delay, $\mathbf{N}_{p}\in \mathbb{C}^{(L_s+T_g)\times L_p}$ is the additive Gaussian noise, $\mathbf{\hat{S}}= [\mathbf{\hat{S}}_1,\mathbf{\hat{S}}_2,\ldots,\mathbf{\hat{S}}_M]\in \mathbb{C}^{(L_s+T_g)\times(M(T_g+1))}$ denotes the expanded spreading matrix which contains $M$ sub-matrices arranged together in columns. The $m$-th $(m=1,\ldots,M)$ sub-matrix is defined as $\mathbf{\hat{S}}_m= [\mathbf{\hat{s}}_{m,0},\mathbf{\hat{s}}_{m,1},\ldots,\mathbf{\hat{s}}_{m,T_g}]\in \mathbb{C}^{(L_s+T_g)\times(T_g+1)}$, which contains $T_g+1$ columns denoting the symbol delays from $0$ to $T_g$, respectively. Thus, the $t$-th column in $m$-th sub-matrix is $\mathbf{\hat{s}}_{m,t}=[\mathbf{0}_{t}^{T},\mathbf{s}_{m}^{T},\mathbf{0}_{T_g-t}^{T}]^{T}$. Then we combine the pilot symbols and channel coefficient to be estimated in a matrix $\mathbf{U}=[\mathbf{U}_{1},\mathbf{U}_{2},\ldots,\mathbf{U}_{M}]^{T}\in \mathbb{C}^{(M(T_g+1))\times L_p}$. For convenience, we refer to it as the pilot matrix and it contains $M$ sub-matrices merged by rows. The $m$-th sub-matrix is defined as $\mathbf{U}_{m}=[\mathbf{u}_{m,1},\mathbf{u}_{m,2},\ldots,\mathbf{u}_{m,T_g+1}] \in \mathbb{C}^{L_p\times(T_g+1 )} $ and refers to the $m$-th spreading sequences. The $t$-th row in $m$-th sub-matrix is $\mathbf{u}_{m,t}=\sum\limits_{k\in\mathcal{K}_{m,t}}h^{k}\mathbf{p}_{k}$, where $\mathcal{K}_{m,t}\subset\mathcal{K}$ contains the index of active users that have selected the sequence $\mathbf{s}_{m}$ and having a delay of $t$. $\mathbf{U}_{m}$ is a zero matrix for non-selected sequence $\mathbf{s}_m$. As only a small number of users are active, $\mathbf{U}_m$ is a row-sparse matrix. By recovering the sparse matrix $\mathbf{U}_m$, we can determine the spreading sequences used by active users and the time delay of active users based on the indexes of the non-zero rows. At the same time, the contention-based access allows the active users to use the same spreading sequence, distinguished by different symbol delays, and therefore can support more users than contention-free access.

The data signal $\mathbf{Y}_{d} \in \mathbb{C}^{(L_s+T_g)\times L_d} $ received at the BS is given by
\begin{equation}
\mathbf{Y}_{d} =\sum\limits_{k\in\mathcal{K}}h^{k}\mathbf{\hat{s}}^{k}_{t_k}\mathbf{d}_{k}^{T}+\mathbf{N}_{d}
=\mathbf{\hat{S}}\mathbf{Z}+\mathbf{N}_{d},
\label{Y_data}
\end{equation}
where $\mathbf{N}_{d}\in \mathbb{C}^{(L_s+T_g)\times L_d}$ is the additive Gaussian noise. According to the same construction, we can represent the received data signal as the product of the extended spreading matrix and the data matrix $\mathbf{Z}=[\mathbf{Z}_{1},\mathbf{Z}_{2},\ldots,\mathbf{Z}_{M}]^{T}\in \mathbb{C}^{(M(T_g+1))\times L_d}$. The data matrix also contains $M$ sub-matrices merged by rows. The $m$-th sub-matrix is defined as $\mathbf{Z}_{m}=[\mathbf{z}_{m,1},\mathbf{z}_{m,2},\ldots,\mathbf{z}_{m,T_g+1}] \in \mathbb{C}^{L_d\times(T_g+1 )} $, and $\mathbf{z}_{m,t}=\sum\limits_{k\in\mathcal{K}_{m,t}}h^{k}\mathbf{d}_{k}$, $\mathcal{K}_{m,t}\subset\mathcal{K}$ contains the index of active users who select sequence $\mathbf{s}_{m}$ and have delay $t$. $\mathbf{Z}_{m}$ is a zero matrix for non-selected sequence $\mathbf{s}_m$.

\subsection{Scenario with Multiple Antennas}

Furthermore, we consider the situation where the BS is equipped with $R$ antennas. In such situations, the channel vector of user $k$ is $\mathbf{h}^{k}=[h^{k,1},h^{k,2},\ldots,h^{k,R}] \in \mathbb{C}^{R}$. Then the received pilot signal $\mathbf{Y}_{p,r}\in \mathbb{C}^{(L_s+T_g)\times L_p}$ at the $r$-th antenna of the BS is given by
\begin{equation}
\mathbf{Y}_{p,r} =\sum\limits_{k\in\mathcal{K}}h^{k,r}\mathbf{\hat{s}}_{t_k}^{k}\mathbf{p}_{k}^{T}+\mathbf{N}_{p,r},
\label{Y_pilot_antenna}
\end{equation}
and the received pilot signal $\mathbf{Y}_{p}\in \mathbb{C}^{(L_s+T_g)\times RL_p}$ at all antenna of the BS can be expressed as
\begin{equation}
\mathbf{Y}_{p}
=\sum\limits_{k\in\mathcal{K}}\mathbf{\hat{s}}_{t_k}^{k}\mathbf{q}_{k}^{T}+\mathbf{N}_{p}
= \mathbf{\hat{S}}\mathbf{U} +\mathbf{N}_{p},
\label{Y_pilot_mmv}
\end{equation}
where $\mathbf{q}_{k}=[h^{k,1}\mathbf{p}_{k}^{T},h^{k,2}\mathbf{p}_{k}^{T},\ldots,h^{k,R}\mathbf{p}_{k}^{T}]^{T}\in \mathbb{C}^{ RL_p\times 1}$, $\mathbf{U}=[\mathbf{U}_{1},\mathbf{U}_{2},\ldots,\mathbf{U}_{M}]^{T}\in \mathbb{C}^{(M(T_g+1))\times RL_p}$ with $\mathbf{U}_{m}=[\mathbf{u}_{m,1},\mathbf{u}_{m,2},\ldots,\mathbf{u}_{m,T_g+1}] \in \mathbb{C}^{RL_p\times(T_g+1 )}$, $\mathbf{u}_{m,t}=\sum\limits_{k\in\mathcal{K}_{m,t}}\mathbf{q}_{k}$,
$\mathcal{K}_{m,t}\subset\mathcal{K}$ contains the index of active users who select sequence $\mathbf{s}_{m}$ and with delay $t$, and $\mathbf{U}_{m}$ is a zero matrix for non-selected sequence $\mathbf{s}_m$.

\par
Similarly, the received data signal $\mathbf{Y}_{d,r} \in \mathbb{C}^{(L_s+T_g)\times L_d}$ at the $r$-th antenna of the BS is given by
\begin{equation}
\mathbf{Y}_{d,r} =\sum\limits_{k\in\mathcal{K}}h^{k,r}\mathbf{\hat{s}}_{k,t_k}\mathbf{d}_{k}^{T}+\mathbf{N}_{d,r},
\label{Y_data_antenna}
\end{equation}
and the received data signal under multiple antennas $\mathbf{Y}_{d} \in \mathbb{C}^{(L_s+T_g)\times RL_d}$ is given by
\begin{equation}
\mathbf{Y}_{d} =\sum\limits_{k\in\mathcal{K}}\mathbf{\hat{s}}_{k,t_k}\mathbf{m}_{k}^{T}+\mathbf{N}_{d}
= \mathbf{\hat{S}}\mathbf{Z} +\mathbf{N}_{d},
\label{Y_data_mmv}
\end{equation}
where $\mathbf{m}_{k}=[h^{k,1}\mathbf{d}_{k}^{T},h^{k,2}\mathbf{d}_{k}^{T},\ldots,h^{k,R}\mathbf{d}_{k}^{T}]^{T}\in \mathbb{C}^{RL_d\times 1 }$, $\mathbf{Z}=[\mathbf{Z}_{1},\mathbf{Z}_{2},\ldots,\mathbf{Z}_{M}]^{T}\in \mathbb{C}^{(M(T_g+1))\times RL_d}$ with $\mathbf{Z}_{m}=[\mathbf{z}_{m,1},\mathbf{z}_{m,2},\ldots,\mathbf{z}_{m,T_g+1}] \in \mathbb{C}^{RL_d\times(T_g+1 )} $, $\mathbf{z}_{m,t}=\sum\limits_{k\in\mathcal{K}_{m,t}}\mathbf{m}_{k}$, $\mathcal{K}_{m,t}\subset\mathcal{K}$ contains the index of active users who select sequence $\mathbf{s}_{m}$ and with delay $t$, and $\mathbf{Z}_{m}$ is a zero matrix for non-selected sequence $\mathbf{s}_m$.

\section{The Proposed Receiver}

Generally, the activity detection problem is constructed as a single measurement vector (SMV) sparse recovery problem. In multiple antennas scenario, the recovered signal is a row-sparse matrix, where the indexes of the non-zero elements are the same for each column. Thus, the MMV problems achieve better recovery performance. Besides, in MMV problems, with more measurements, the sparse recovery performance becomes better. In this paper, we consider the sparsity consistency between the pilot matrix $\mathbf{U}$ and the data matrix $\mathbf{Z}$ according to the joint transmission of pilot and data, and construct the joint channel estimation and data recovery in a single-antenna scenario as an MMV problem. We also extend the model to multiple antennas, where the corresponding problem becomes MMV problems with more columns. Compared to other papers that consider multiple antennas, the proposed receiver solves an MMV problem with more columns (pilot + data), which enables improved detection performance. Furthermore, we introduce the special sparse structure in asynchronous access with data length diversity, which can be explored at the receiver for performance enhancement. Then, we propose a novel neural network based on deep-unfolding LAMP to explore the multiple types of sparsity.

\subsection{Problem Formulation}

To solve the activity detection and data recovery jointly, we combine the received pilot signal and data signal into one equation as
\begin{equation}
\mathbf{Y}=[\mathbf{Y}_{p},\mathbf{Y}_{d}]
=\mathbf{\hat{S}}[\mathbf{U},\mathbf{Z}]+\mathbf{N}_{c},
=\mathbf{\hat{S}}\mathbf{X}+\mathbf{N}_{c},
\label{Y_combine}
\end{equation}
where $\mathbf{Y} \in \mathbb{C}^{(L_s+T_g)\times R(L_p+L_d)}$ is obtained by splicing the received pilot signal $\mathbf{Y}_{p}$ and data signal $\mathbf{Y}_{d}$ in columns, $\mathbf{X} \in \mathbb{C}^{N(T_g+1)\times R(L_p+L_d)}$ ($R=1$ for single antenna scenario) is obtained by splicing the pilot matrix $\mathbf{U}$ and data matrix $\mathbf{Z}$ in columns. To simplify, we define $\tilde{M}=L_s+T_g, \tilde{N}=N(T_g+1)$ and $L=L_p+L_d$. Since the pilot matrix $\mathbf{U}$ and the data matrix $\mathbf{Z}$ have the same non-zero rows, combining them as a sparse matrix $\mathbf{X}$ with more columns for recovery leads to a higher performance gain as compared to solving them separately.

This MMV optimization problem can be formulated as:
\begin{equation}
\min_{\mathbf{X}} \,  \|\mathbf{X}\| _{\text{row},0}  \ \
\text{s.t.}\ \
\| \mathbf{Y}-\mathbf{\hat{S}}\mathbf{X}\| _\mathbf{F}^2 \leq \varepsilon,
\label{opti_Y_pilot_mmv}
\end{equation}
where $\|\cdot\| _{\text{row},0}$ counts the number of non-zeros rows of a matrix, and $\varepsilon > 0$ that depends on the signal-to-noise ratio (SNR). While there are many approaches to address (\ref{opti_Y_pilot_mmv}), in asynchronous access with data length diversity we take advantage of the fact that $\mathbf{X}$ is a matrix with structural sparsity and propose a detection algorithm that outperforms the existing algorithms.

\begin{figure}[!t]%
\centering%
\includegraphics[width=0.48\textwidth]{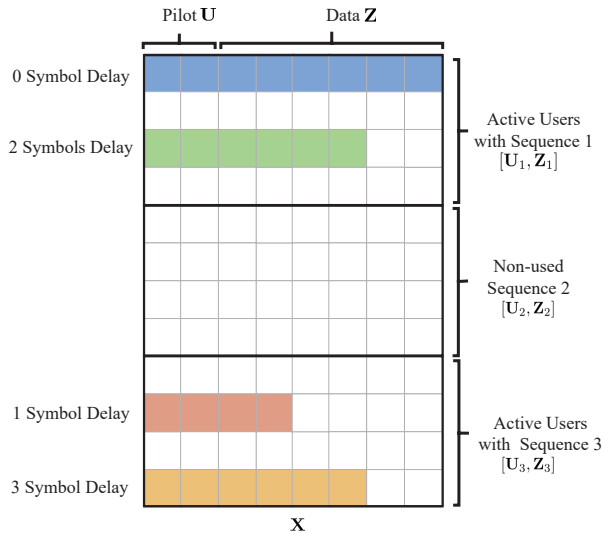}%
\DeclareGraphicsExtensions. \caption{An illustration of structured sparsity in asynchronous massive access with data length diversity with the setting of $M=3, L_p=2$, and $L_d=6$. } \label{fig:multi_sparse}
\end{figure}%

In asynchronous access with data length diversity, $\mathbf{X}$ is a matrix with three types of sparsity, as shown in Fig.~\ref{fig:multi_sparse}. The first one is, intuitively, the row-sparsity due to user activity, which depends on the number of active users that select different pilots or have different symbol-level delay. The second type of sparsity is the block sparsity. We can divide $\mathbf{X}$ into $M$ blocks, each of which contains $T_g+1$ rows. Define the index set of selected pilots as $\mathcal{M}$, and only the blocks in $\mathcal{M}$ have non-zeros rows. The third type of sparsity comes from the data length diversity. As different devices transmit different number of data symbols, some non-zero rows of $\mathbf{X}$ are sparse and have many zeros in the end of the rows. Thus, we can enhance the detection algorithm by exploring the different types of sparse structures in $\mathbf{X}$. Besides, compared to contention-free CS-based asynchronous access, our contention-based access can use the symbol asynchronous information to distinguish active users that selected the same sequences, thereby increasing the access performance.

\subsection{Model-Driven AMP Receiver}

Firstly, we introduce the general AMP algorithm for MMV problems. The $t$-th iteration of AMP-MMV is given by
\begin{subequations}
\begin{align}
& \mathbf{V}^{t}=\mathbf{Y}-\mathbf{\hat{S}}\mathbf{\hat{X}}^{t}+ b^t \mathbf{V}^{t-1},\\
& \mathbf{\hat{X}}^{t+1}=\eta_{vst}(\mathbf{\hat{X}}^{t}+\mathbf{\hat{S}}^{T}\mathbf{V}^{t};\lambda^{t}),
\end{align}
\label{itr_amp_step}
\end{subequations}
where $\mathbf{V}^{-1}=\mathbf{0}, \mathbf{\hat{X}}^{0}=\mathbf{0}$, and
\begin{equation}
b^t=\frac{1}{\tilde{M}RL}\|\mathbf{\hat{X}}^{t}\| _{0},
\label{para1_amp_mmv}
\end{equation}

\begin{equation}
\lambda^{t}=\frac{\alpha^{t}}{\sqrt{\tilde{M}RL}}\|\mathbf{V}^{t}\| _{2},
\label{para2_amp_mmv}
\end{equation}
$\alpha$ is a tuning parameter, $\eta_{vst}(\mathbf{x};\lambda)$ is a row-thresholding function operates on each row as
\begin{equation}
\eta_{vst}(\mathbf{x};\lambda)=\mathbf{x}\ast max\{1-\frac{\lambda}
{\|\mathbf{x}\|_{2}},0 \}.
\label{rowthres_amp_mmv}
\end{equation}

Define $\mathbf{X}=[\mathbf{X}_{1},\mathbf{X}_{2},\cdots,\mathbf{X}_{L}]$. According to the data length diversity, as shown in Fig.~\ref{fig:multi_sparse}, the last few columns of $\mathbf{X}$ are sparser. To explore the data length diversity, Xiao et al. propose a backward activity level estimation algorithm in \cite{9268113}. In this paper, we consider the asynchronous access scenario and modify the AMP-MMV algorithm to exploit the data length diversity.

In the modified AMP-MMV with backward estimation (AMP-BP), we divide $\mathbf{X}$ into $L$ sub-matrices with $R$ columns. Those sub-matrices have different sparsity levels due to the data length diversity, while $R$ columns in each sub-matrix contain the signal in $R$ antennas and have the same sparsity level. Next, we estimate the last $R$ columns of $\mathbf{X}$, defined as $\mathbf{X}(:,\mathcal{I}_{L}),\mathcal{I}_{L}=\{R(L-1)+1:RL\}$. As these $R$ columns have minimum sparsity, we can reconstruct them with higher accuracy, and their sparse supports can be exploited to estimate former columns as prior information. With initialization $\mathbf{V}(:,\mathcal{I}_{L})^{-1}=\mathbf{0}$ and $\mathbf{X}(:,\mathcal{I}_{L})^{0}=\mathbf{0}$, the $t$-th iteration of AMP-BP for the estimation of $\mathbf{X}(:,\mathcal{I}_{L})$ is given by
\begin{subequations}
\begin{align}
& \mathbf{V}(:,\mathcal{I}_{L}))^{t}=\mathbf{Y}(:,\mathcal{I}_{L}))
-\mathbf{\hat{S}}\mathbf{\hat{X}}(:,\mathcal{I}_{L}))^{t}
+ \mathbf{V}(:,(\mathcal{I}_{L}))^{t-1},
\label{itr_ampbp_step1} \\
& \mathbf{\hat{X}}(:,\mathcal{I}_{L}))^{t+1}
=\eta_{vst}(\mathbf{\hat{X}}(:,\mathcal{I}_{L}))^{t}
+\mathbf{\hat{S}}^{T}\mathbf{V}(:,\mathcal{I}_{L}))^{t};\lambda^{t}_{L}). \label{itr_ampbp_step2}
\end{align}
\end{subequations}
where
\begin{equation}
b^t_{L}=\frac{1}{\tilde{M}R}\|\mathbf{\hat{X}}(:,\mathcal{I}_{L}))^{t}\| _{0},
\label{para1_amp_bp}
\end{equation}
\begin{equation}
\lambda^{t}_{L}=\frac{\alpha_{L}^{t}}{\sqrt{\tilde{M}R}}\|\mathbf{V}(:,\mathcal{I}_{L}))^{t}\| _{2}.
\label{para2_amp_bp}
\end{equation}

For the remaining columns, we use the result of previous columns as prior information to enhance the sparse signal reconstruction performance. For the estimation of the $i$-th sub-matrix, $i = L-1 ,\cdots, 1 $, we first obtain the index of non-zero rows of estimated columns via
\begin{equation}
\mathcal{S}_{i}= \{ j, \|\mathbf{\hat{X}}(j,\mathcal{I}_{i+1})\|_{2} < \delta \},
\label{pri}
\end{equation}
where $\mathcal{I}_{i+1}=\{Ri+1:RL\}$ and $\mathbf{\hat{X}}(j,\mathcal{I}_{i+1})$ is the $j$-th row of last $R(L-i)$ columns of the estimation matrix
$\mathbf{\hat{X}}(:,\mathcal{I}_{i+1})$ and $\delta$ is a pre-defined parameter.

Then, we initialize the last $\mathcal{I}_{i}=\{R(i-1)+1:RL\}$ columns with
\begin{equation}
\mathbf{\hat{X}}(\mathcal{S}_{i},\mathcal{I}_{i})=\mathbf{\hat{S}}(:,\mathcal{S}_{i})^{-1}\ast\mathbf{Y}_{com}(:,\mathcal{I}_{i}),
\end{equation}
and $\mathbf{V}(:,\mathcal{I}_{i})^{-1}=\mathbf{0}$. The $t$-th iteration for the estimation of $\mathbf{X}(:,\mathcal{I}_{i})$ is given by
\begin{subequations}
\begin{align}
&\mathbf{V}(:,\mathcal{I}_{i})^{t}=\mathbf{Y}(:,\mathcal{I}_{i})-\mathbf{\hat{S}}\mathbf{\hat{X}}(:,\mathcal{I}_{i})^{t} + b^t_{i}\mathbf{V}(:,\mathcal{I}_{i})^{t-1},
\label{itr_ampbp_step3} \\
&\mathbf{\hat{X}}(:,\mathcal{I}_{i})^{t+1} =\eta_{vst_{pr}}(\mathbf{\hat{X}}(:,\mathcal{I}_{i})^{t}
+\mathbf{\hat{S}}^{T}\mathbf{V}(:,\mathcal{I}_{i})^{t};\lambda^{t}_{i};\mathbf{s}_i), \label{itr_ampbp_step4}
\end{align}
\end{subequations}
where
\begin{equation}
b^t_{i}=\frac{1}{\tilde{M}R(L-i+1)}\|\mathbf{\hat{X}}(:,\mathcal{I}_{i}))^{t}\| _{0},
\label{para1_amp_bp}
\end{equation}
\begin{equation}
\lambda^{t}_{i}=\frac{\alpha_{i}^{t}}{\sqrt{\tilde{M}R(L-i+1)}}\|\mathbf{V}(:,\mathcal{I}_{i}))^{t}\| _{2},
\label{para2_amp_bp}
\end{equation}
$\mathbf{s}_i$ is a one-hot vector with $\mathbf{s}_i(\mathcal{S}_{i})=1$ that indicates the prior supports obtained via the estimation of previous columns, $\eta_{vst_{pr}}(\cdot)$ is a thresholding function aided by a prior information and defined as:
\begin{equation}
\eta_{vst_{pr}}(\mathbf{x};\lambda;\mathbf{s})=\eta_{vst}(\mathbf{x};\lambda)-\eta_{vst}(\mathbf{x};\lambda).\ast \mathbf{s}+\mathbf{x}.\ast \mathbf{s}.
\label{act_pri}
\end{equation}
Using this function,  we keep the rows corresponding to $\mathcal{S}$ unchanged and feed the other rows to the row-thresholding function $\eta_{vst}(\cdot)$.

\subsection{The Proposed Data-Driven Receiver}

\begin{figure}[!t]%
\centering%
\includegraphics[width=0.48\textwidth]{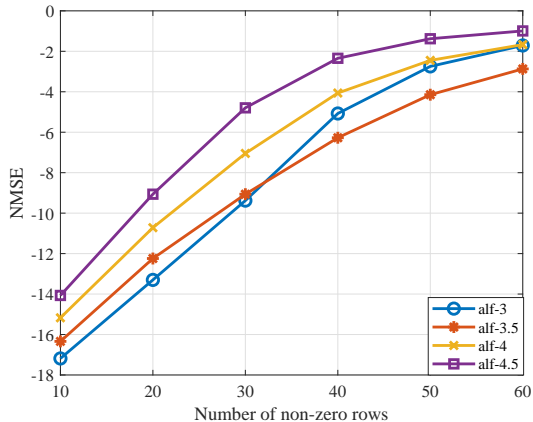}%
\DeclareGraphicsExtensions. \caption{The sparse recovery performance os AMP-MMV with different $\alpha$. The dimension of the sensing matrix is $200\times500$, the number of measurements is $4$, and SNR is $10 dB$. } \label{fig:amp_diff_alf}
\end{figure}%

By exploring the data length diversity, the proposed model-driven AMP-BP algorithm is expected to achieve better performance than the original AMP-MMV algorithm. However, the performance is affected by the thresholding parameter $\alpha$, which is set via cross validation. As shown in Fig.~\ref{fig:amp_diff_alf}, unsuitable parameter leads to performance degradation. In the AMP-BP algorithm, the number of tuned parameters is $L$. This inspires the use of data-driven methods, which learn the parameters from the training data in an end-to-end manner. As an evidence, the data-driven methods, learned AMP (LAMP) \cite{7934066} shows improved performance as compared to the AMP.

Unlike AMP, the LAMP learns the mapping for plenty of training pairs $\{ \mathbf{Y}^{q}, \mathbf{X}^{q}\}_{q=1}^{Q}$ by learning network parameters $\Theta$. Generally, AMP is used to initialize the network parameters which are then updated by back-propagation to minimize the loss function, e.g., the mean square error given by
\begin{equation}
\mathcal{L}(\Theta)= \sum_{q=1}^Q \|\mathbf{\hat{X}}^{q}-\mathbf{X}^{q}\|_{2}^{2}=\|r(\mathbf{Y}^{q}, \Theta)-\mathbf{X}^{q}\|_{2}^{2},
\label{loss_mse}
\end{equation}
where $\mathbf{\hat{X}}$ is the output of the LAMP and $r(\cdot)$ denotes the operations of the LAMP.

As the LAMP in \cite{7934066} is proposed for SMV problems, we unfold the AMP-MMV into a neural network and obtain LAMP-MMV. In tied LAMP-MMV that all layers share the same weight matrix $\mathbf{B}$, the $t$-th layer, which corresponds to the update of AMP-MMV in \eqref{itr_amp_step}, is given by
\begin{subequations}
\begin{align}
& \mathbf{V}^{t}=\mathbf{Y}-\mathbf{\hat{S}}\mathbf{\hat{X}}^{t}+ \frac{1}{\tilde{M}RL}\|\mathbf{\hat{X}}^{t}\| _{0} \mathbf{V}^{t-1},
\label{lamp_step1} \\
& \mathbf{\hat{X}}^{t+1}=\eta_{vst}(\mathbf{\hat{X}}^{t}+\mathbf{B}\mathbf{V}^{t};\frac{\alpha^{t}}{\sqrt{\tilde{M}RL}}\|\mathbf{V}^{t}\| _{2}),
\label{lamp_step2}
\end{align}
\end{subequations}
The learned parameters $\Theta$ of LAMP-MMV is $\Theta =\{\mathbf{B}, \{\alpha^{t}\}_{t=1}^{T} \}$.

\begin{figure*}[!t]%
\centering%
\includegraphics[width=0.8\textwidth]{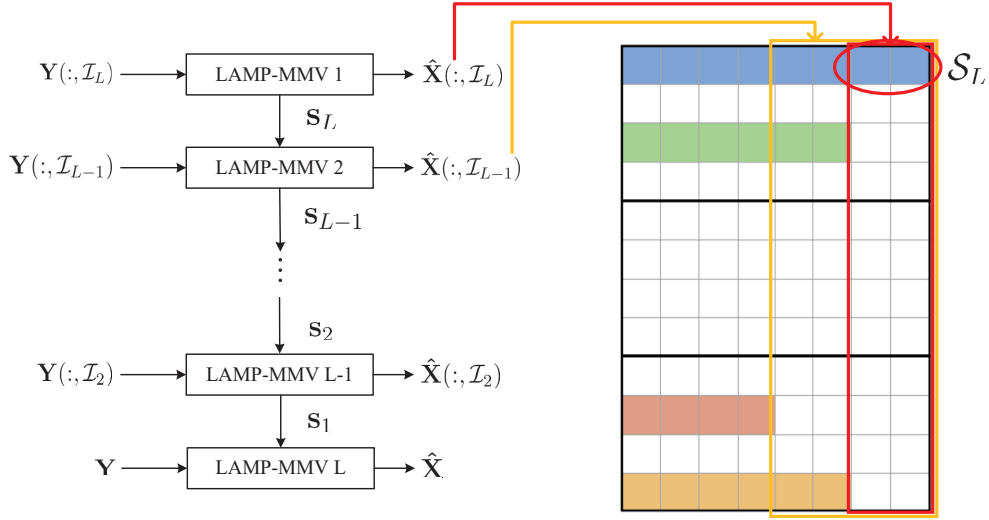}%
\DeclareGraphicsExtensions. \caption{The main structure of the proposed LAMP-BP, which estimate the structural sparse signal via backward estimation. } \label{fig:lamp_bp1}
\end{figure*}%

\begin{figure}[!t]%
\centering%
\includegraphics[width=0.48\textwidth]{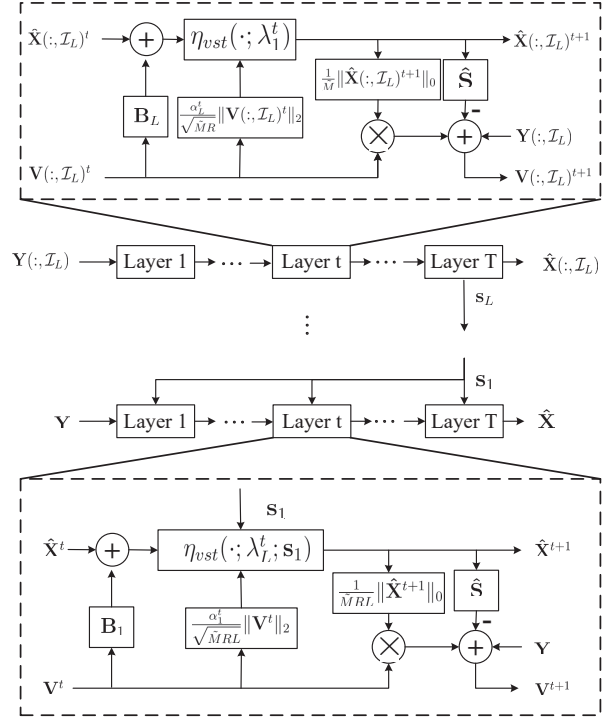}%
\DeclareGraphicsExtensions. \caption{The detailed structure in each layer of the proposed LAMP-BP with tied weights.} \label{fig:lamp_bp2}
\end{figure}%

In this paper, we extend the AMP-BP into a neural network following the same way of deep unfolding, and the resulting method is named LAMP-BP. As shown in Fig.~\ref{fig:lamp_bp1}, the whole network contains $L$ subnetworks. Each subnetwork deals with more columns of $\mathbf{Y}$ than the former one. Of course, the later sub-networks need to estimate a matrix that is less sparse due to the data length diversity. In addition to the first subnetwork, the other subnetworks have the aid of prior information obtained from the preceding subnetwork. The $t$-th layer of the $i$-th ($i=L,\cdots,1$) subnetwork is given by
\begin{subequations}
\begin{align}
&\mathbf{V}(:,\mathcal{I}_{i})^{t}=\mathbf{Y}(:,\mathcal{I}_{i})-\mathbf{\hat{S}}\mathbf{\hat{X}}(:,\mathcal{I}_{i})^{t} + b^t_{i}\mathbf{V}(:,\mathcal{I}_{i})^{t-1},
\label{itr_lampbp_step3} \\
&\mathbf{\hat{X}}(:,\mathcal{I}_{i})^{t+1} =\eta_{vst_{pr}}(\mathbf{\hat{X}}(:,\mathcal{I}_{i})^{t}
+\mathbf{B}\mathbf{V}(:,\mathcal{I}_{i})^{t};\lambda^{t}_{i};\mathbf{s}_i), \label{itr_lampbp_step4}
\end{align}
\end{subequations}
where
\begin{equation}
b^t_{i}=\frac{1}{\tilde{M}R(L-i+1)}\|\mathbf{\hat{X}}(:,\mathcal{I}_{i}))^{t}\| _{0},
\label{para1_lamp_bp}
\end{equation}
\begin{equation}
\lambda^{t}_{i}=\frac{\alpha_{i}^{t}}{\sqrt{\tilde{M}R(L-i+1)}}\|\mathbf{V}(:,\mathcal{I}_{i}))^{t}\| _{2},
\label{para2_lamp_bp}
\end{equation}
and $\mathbf{s}_{L}=\mathbf{0}$ and $\mathbf{s}_{i} (i=L-1,\cdots,1)$ are obtained according to the equation \eqref{pri}. Fig.~\ref{fig:lamp_bp2} shows the detailed structure of each layer in the LAMP-BP. In each subnetwork, the learned parameters are $\Theta_{i} =\{\mathbf{B}_{i}, \{\alpha^{t}_{i}\}_{t=1}^{T} \} (i= L,\cdots, 1)$. Different subnetworks deal with the MMV problems with different sparsity. The thresholding parameters are different for each sub-network, while the weight parameters $\mathbf{B}_{i}$ can be shared.

In the training phase, we firstly train the last sub-network with the following loss function
\begin{equation}
\begin{split}
\mathcal{L}(\Theta)&= \sum_{q=1}^Q \|\mathbf{\hat{X}}(:,\mathcal{I}_{L}))^{q}-\mathbf{X}(:,\mathcal{I}_{L}))^{q}\|_{2}^{2}\\
&=\|r_{L}(\mathbf{Y}(:,\mathcal{I}_{L}))^{q}, \Theta)-\mathbf{X}(:,\mathcal{I}_{L}))^{q}\|_{2}^{2}.
\label{loss_subnet1}
\end{split}
\end{equation}
We fix the parameters of this subnetwork and train the next subnetwork with a loss function that involves more columns of $\mathbf{X}$
\begin{equation}
\begin{split}
\mathcal{L}(\Theta)&= \sum_{q=1}^Q \|\mathbf{\hat{X}}(:,\mathcal{I}_{L-1}))^{q}-\mathbf{X}(:,\mathcal{I}_{L-1}))^{q}\|_{2}^{2}\\
&=\|r_{L-1}(\mathbf{Y}(:,\mathcal{I}_{L-1}))^{q}, \Theta)-\mathbf{X}(:,\mathcal{I}_{L-1}))^{q}\|_{2}^{2}.
\end{split}
\end{equation}

\subsection{Theoretical Analysis}

Here we analyze the performance gain of the proposed algorithm. The sparsity of $\mathbf{X}$ is directly related to the number of active devices. We prove that the proposed algorithm is able to support more active devices by recovering a  sparser signal. According to \cite{1453780}, the maximum sparsity that can be solved in the noiseless MMV problem is given by Lemma 1.

\begin{lemma}
[\textbf{Lemma 1 from \cite{1453780}}]
For a noiseless MMV problem $\mathbf{Y} \in \mathbb{C}^{\tilde{M} \times \tilde{L}}
=\mathbf{\hat{S}}\mathbf{X}$, $\mathbf{\hat{S}} \in \mathbb{C}^{\tilde{M} \times \tilde{N}}$. Any $\tilde{M}$ columns of $\mathbf{\hat{S}}$ are linearly independent and $rank(\mathbf{Y})= \tilde{L} \leq \tilde{M}$. There exist a unique sparse solution $\mathbf{\hat{X}}$ with sparsity $r$ and $r\leq \lceil (\tilde{M}+\tilde{L})/2\rceil -1 $, where $\lceil\cdot\rceil$ is the ceiling operation.
\end{lemma}

 Theorem 1 below states that the maximum sparsity allowed in the noiseless MMV problem is increased with the knowledge of the index of partial non-zero elements.

\begin{theorem}
[\textbf{Sufficient Condition}]
For a noiseless MMV problem $\mathbf{Y} \in \mathbb{C}^{\tilde{M} \times \tilde{L}}
=\mathbf{\hat{S}}\mathbf{X}$, $\mathbf{\hat{S}} \in \mathbb{C}^{\tilde{M} \times \tilde{N}}$. Any $\tilde{M}$ columns of $\mathbf{\hat{S}}$ are linearly independent and $rank(\mathbf{Y})= \tilde{L} \leq \tilde{M}$. With known $r_s$ non-zero index of $\mathbf{X}$, there exist a unique sparse solution $\mathbf{\hat{X}}$ with sparsity $r$ and $r\leq \lceil (\tilde{M}+\tilde{L}+r_s)/2\rceil -1 $, where $\lceil\cdot\rceil$ is the ceiling operation.
\end{theorem}

Proof: We show that all other sparse solutions have sparsity larger than $r$. We assume that $\mathbf{X}_{1}$ and $\mathbf{X}_{2}$ are two sparse solutions of $\mathbf{Y}=\mathbf{\hat{S}}\mathbf{X}$. The sparsity levels of $\mathbf{X}_{1}$ and $\mathbf{X}_{2}$ are $r_1$ and $r_2$, respectively.
$\mathbf{X}_{1}^{r_1}$ and $\mathbf{X}_{2}^{r_2}$ represent sub-matrices that contain the nonzero rows of $\mathbf{X}_{1}$ and $\mathbf{X}_{2}$, respectively. Then, we have
\begin{equation}
  \begin{bmatrix}
  \mathbf{\hat{S}}^{r_1} \ \mathbf{\hat{S}}^{r_2}
  \end{bmatrix}
  *
  \begin{bmatrix}
  \mathbf{X}_{1}^{r_1}\\ -\mathbf{X}_{2}^{r_2}
  \end{bmatrix}=0,
\end{equation}
where $\mathbf{\hat{S}}^{r_1}$ and $\mathbf{\hat{S}}^{r_2}$ have $r_1$ and $r_2$ columns, respectively. As a prior information, we have the known non-zero indices of $r_s$. Thus both the matrix $\mathbf{\hat{S}}^{r_1}$ and $\mathbf{\hat{S}}^{r_2}$ contain $r_s$ columns in the matrix $\mathbf{\hat{S}}$ corresponding to these non-zero rows. With the assumption that any $\tilde{M}$ columns of $\mathbf{\hat{S}}$ are linearly independent and $rank(\mathbf{Y})= \tilde{L} \leq \tilde{M}$, the matrix $ [\mathbf{\hat{S}}^{r_1},\mathbf{\hat{S}}^{r_2}]$ has a null space of dimension at least $\tilde{L}$. This means $r_1+r_2-r_s\geq \tilde{M}+\tilde{L}$. If $r_1\leq \lceil (\tilde{M}+\tilde{L}+r_s)/2\rceil -1 $, and then we have $r_2>r_1$. As all other sparse solutions have higher sparsity than $\lceil (\tilde{M}+\tilde{L}+r_s)/2\rceil -1 $, there exist a unique sparse solution with sparsity $r$ and $r\leq \lceil (\tilde{M}+\tilde{L}+r_s)/2\rceil -1 $.

According to this Theorem, we can recovery less sparse signals by assuming the partial knowledge about the non-zero indices. This means that, compared with general massive access scenario without data length diversity, the structure in our model enables to support a larger number of active users.

\section{Numerical Results}

In this section, we construct various numeral experiments to show the performance gain of the proposed algorithm under different settings. We consider an OFDM system with 1.4 MHz bandwidth and 72 subcarriers in total. For channel coding, we use a (2,1,6) convolutional code and a (4,6) RS code. The modulation type is 16 QAM. The spreading sequences are generated via Gaussian distribution and then normalized column by column. The number of spreading sequences is 100, and the length is 70. The number of transmitted pilot symbols of all users is 1. The number of transmitted data symbols follows the discrete uniform distribution in the range of $[1,3]$. The maximum number of symbol delays and guard time is set to 3. The SNR is set to 30 dB. For each network, we generate 2 million training pairs. The batchsize and the initial learning rate are 1000 and 0.1, respectively. Every time the loss function stops decreasing or the number of iterations reaches the maximum number 300,000, the learning rate is decayed by multiplying it by 0.1 until it decreases to 0.0001.
 
We use $F$ measure to jointly measure the ratio of misdetection and erroneous pilot detection, defined as
\begin{equation}
\begin{split}
F_{1measure}= \frac{2\mu_{p}\mu_{r}}{\mu_{p}+\mu_{r}},
\label{NMSE}
\end{split}
\end{equation}
where $\mu_{p}=\frac{card(\mathcal{M}\bigcap\hat{\mathcal{M}})}{card(\mathcal{M})}$ defines the precision rate of selected pilots, $\mu_{r}=\frac{card(\mathcal{M}\bigcap\hat{\mathcal{M}})}{card(\hat{\mathcal{M}})}$ defines the recall rate, $\hat{\mathcal{M}}$ denotes the index of detected pilots, and $card(\cdot)$ denotes the number of elements in the set. Compared with the successful detection rate of selected pilots $\mu_{p}$, $F_{1measure}$ is more comprehensive by jointly considering the successful detection rate and the false alarm rate. For channel estimation performance, it is hard to compare the NMSE of channel of each user as different users may select the same pilot. We evaluate the normalized mean square error (NMSE) (in dB) of $\mathbf{U}$, which is given by
\begin{equation}
\begin{split}
NMSE (dB)= 10 \log_{10} (\frac{\| \mathbf{\hat{U}}-\mathbf{U} \|_{F}^{2}}{\|\mathbf{U} \|_{F}^{2}}).
\label{NMSE}
\end{split}
\end{equation}
For data recovery performance, we compare the ratio of active users that correctly recovered their data, which is defined as
\begin{equation}
\begin{split}
\mu_{data}= \frac{card(\mathcal{K}\bigcap\hat{\mathcal{K}})}{N_{a}},
\label{mu_data}
\end{split}
\end{equation}
where $\hat{\mathcal{K}}$ denotes the index of active users that successfully decoded their transmitted data.

\begin{figure}
\centering
\subfigure[]{
\includegraphics[width=0.48\textwidth]{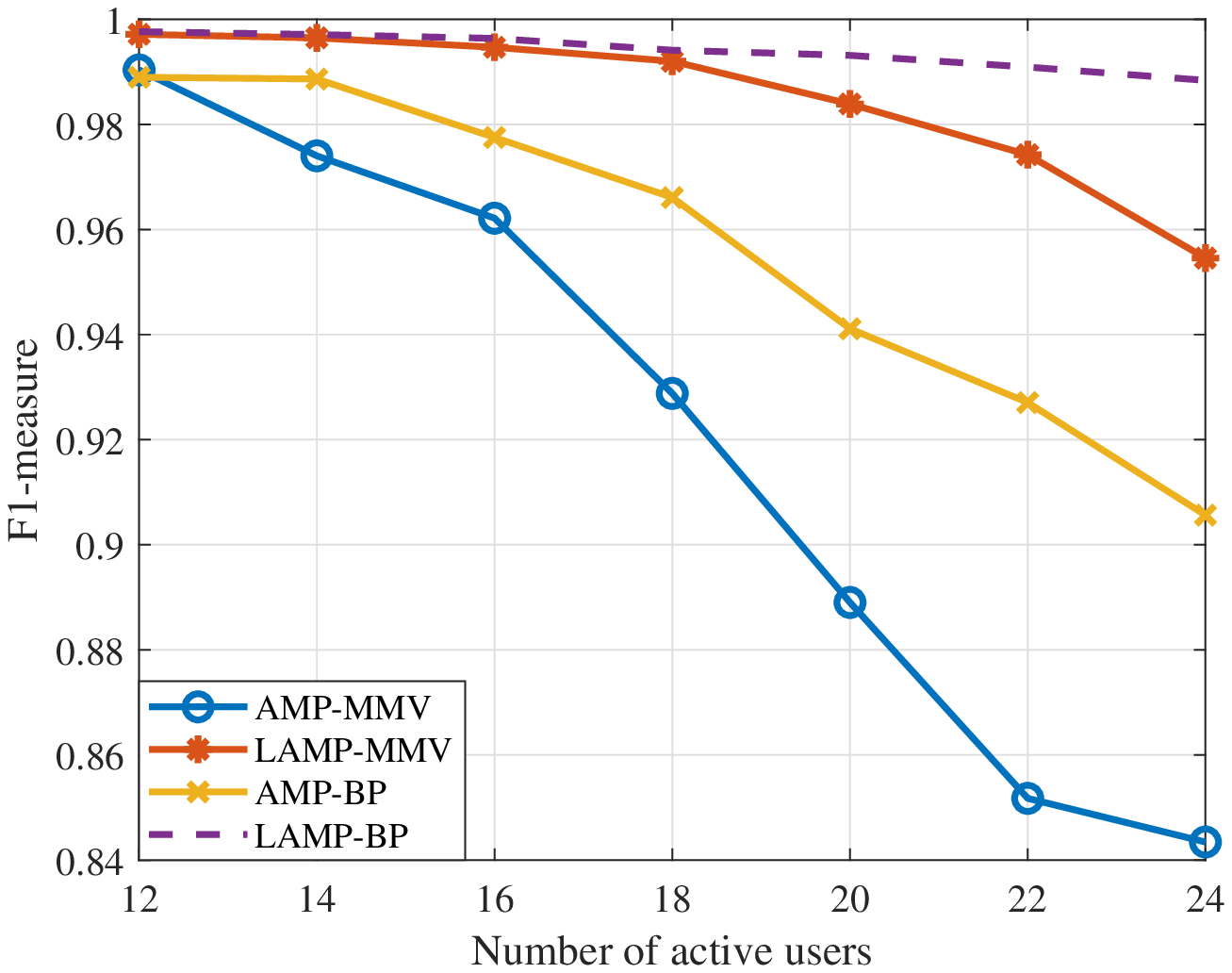}}
\vspace{.1in}
\subfigure[]{
\includegraphics[width=0.48\textwidth]{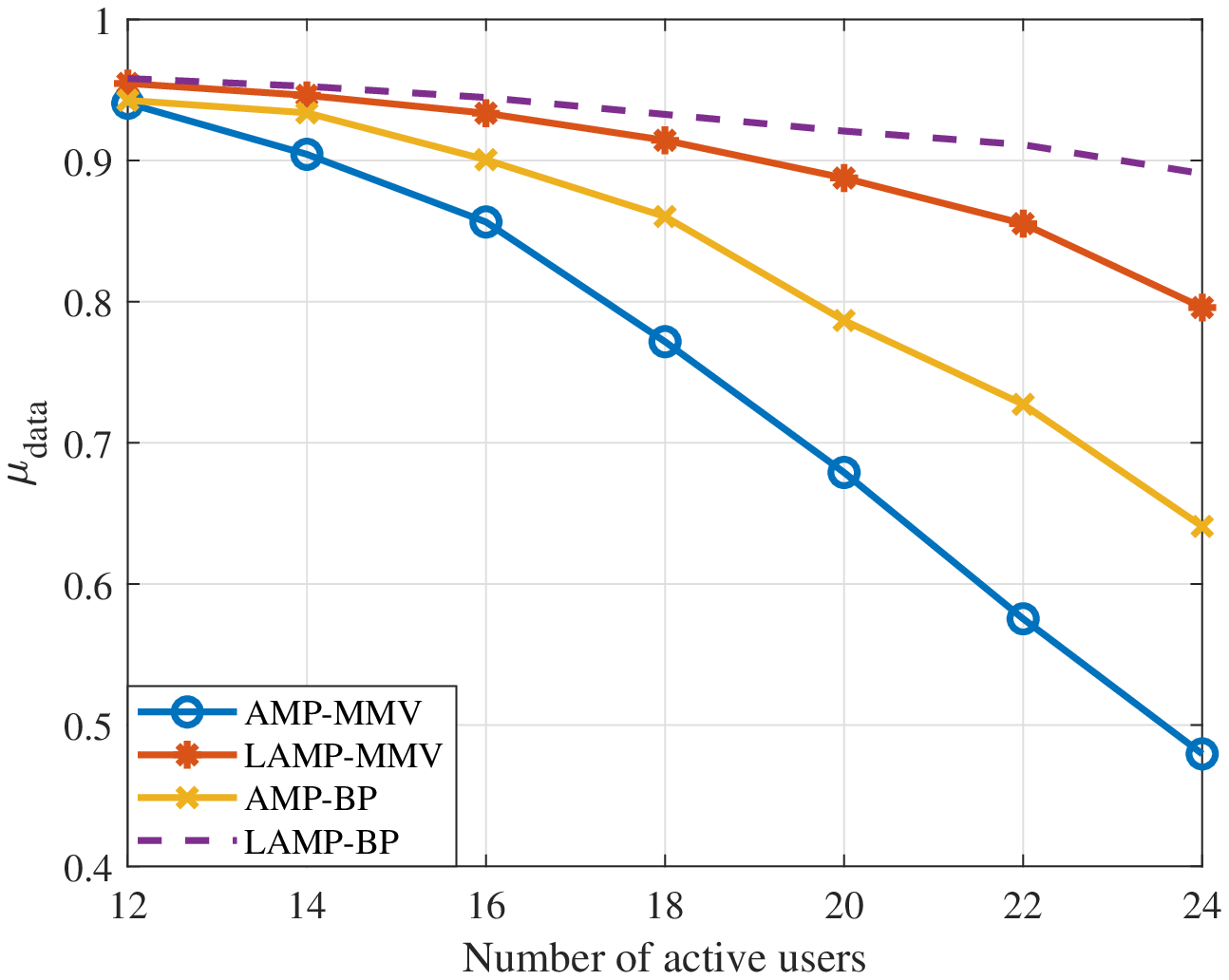}}
\caption{Pilot detection and data recovery performance under different number of active users, where $L_s=70$, $L_p=1$, $L_d=3$, $T_g=3$, $R=1$, SNR$=30dB$.}
\label{fig:diffact} 
\end{figure}

\begin{figure}
\centering
\includegraphics[width=0.48\textwidth]{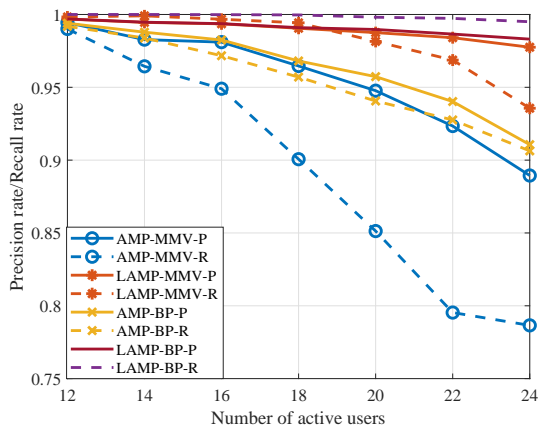}
\caption{Precision and recall of pilot detection under different number of active users, where $L_s=70$, $L_p=1$, $L_d=3$, $T_g=3$, $R=1$, SNR$=30dB$.}
\label{fig:diffact1} 
\end{figure}

In Fig.~\ref{fig:diffact}, we compare the performance of pilot detection and data recovery under different numbers of active users. With the increase of the number of active user, the joint pilot detection and data recovery under higher sparsity becomes more difficult. The proposed AMP-BP and LAMP-BP show improved performance than AMP and LAMP respectively in both pilot detection and data recovery. The proposed LAMP-BP has the best performance than the CS algorithm AMP and data-driven method LAMP. To further analyze the performance gain, we show the pilot detection correct ratio (with label -P) and false alarm ratio (with label -R) in Fig.~\ref{fig:diffact1}. It shows that the proposed method greatly decreases the false alarm ratio and thus increases the accuracy of channel estimation. With improved performance of channel estimation, we can obtain a higher data recovery rate.

\begin{figure}
\centering
\subfigure[]{
\includegraphics[width=0.48\textwidth]{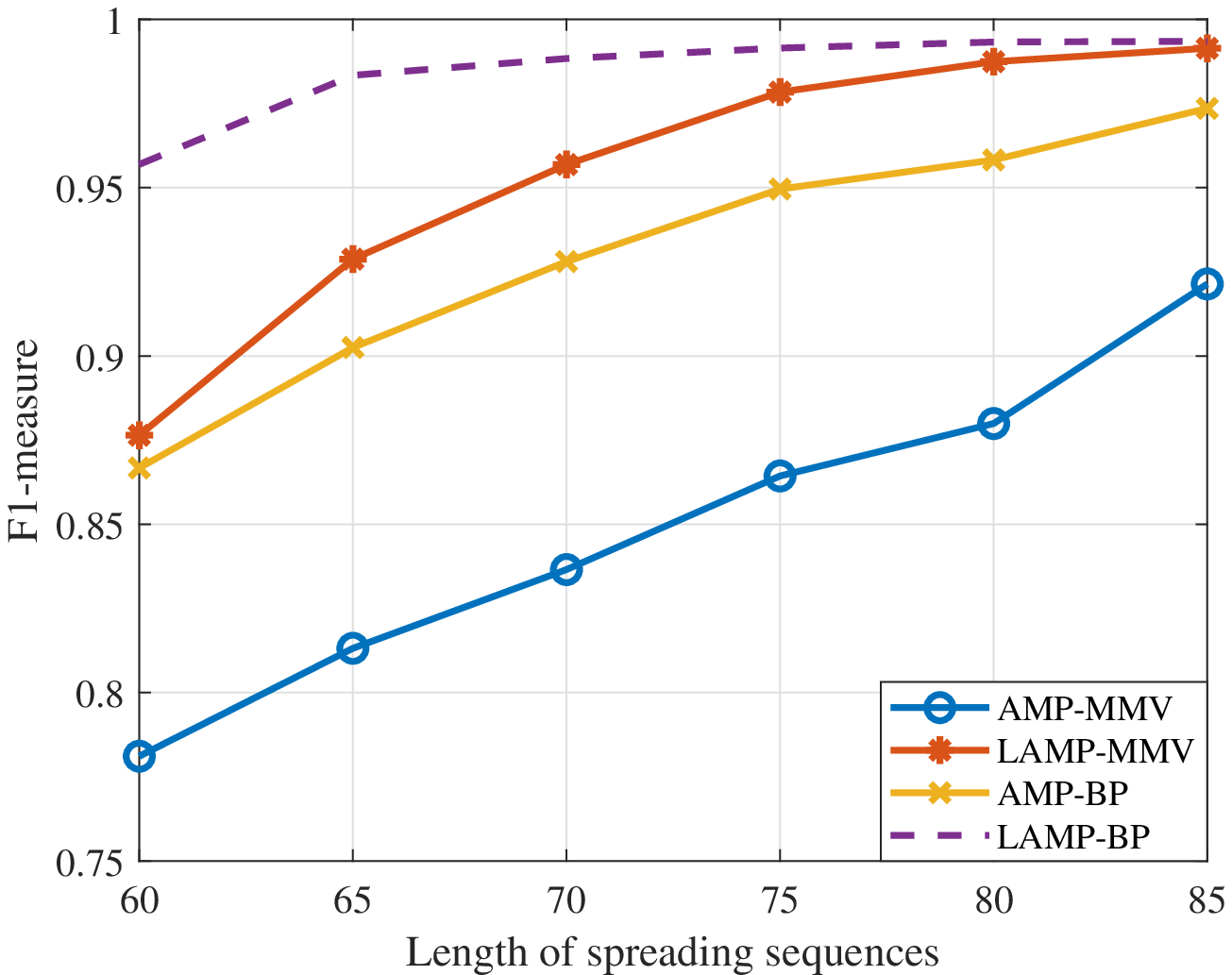}}
\vspace{.1in}
\subfigure[]{
\includegraphics[width=0.48\textwidth]{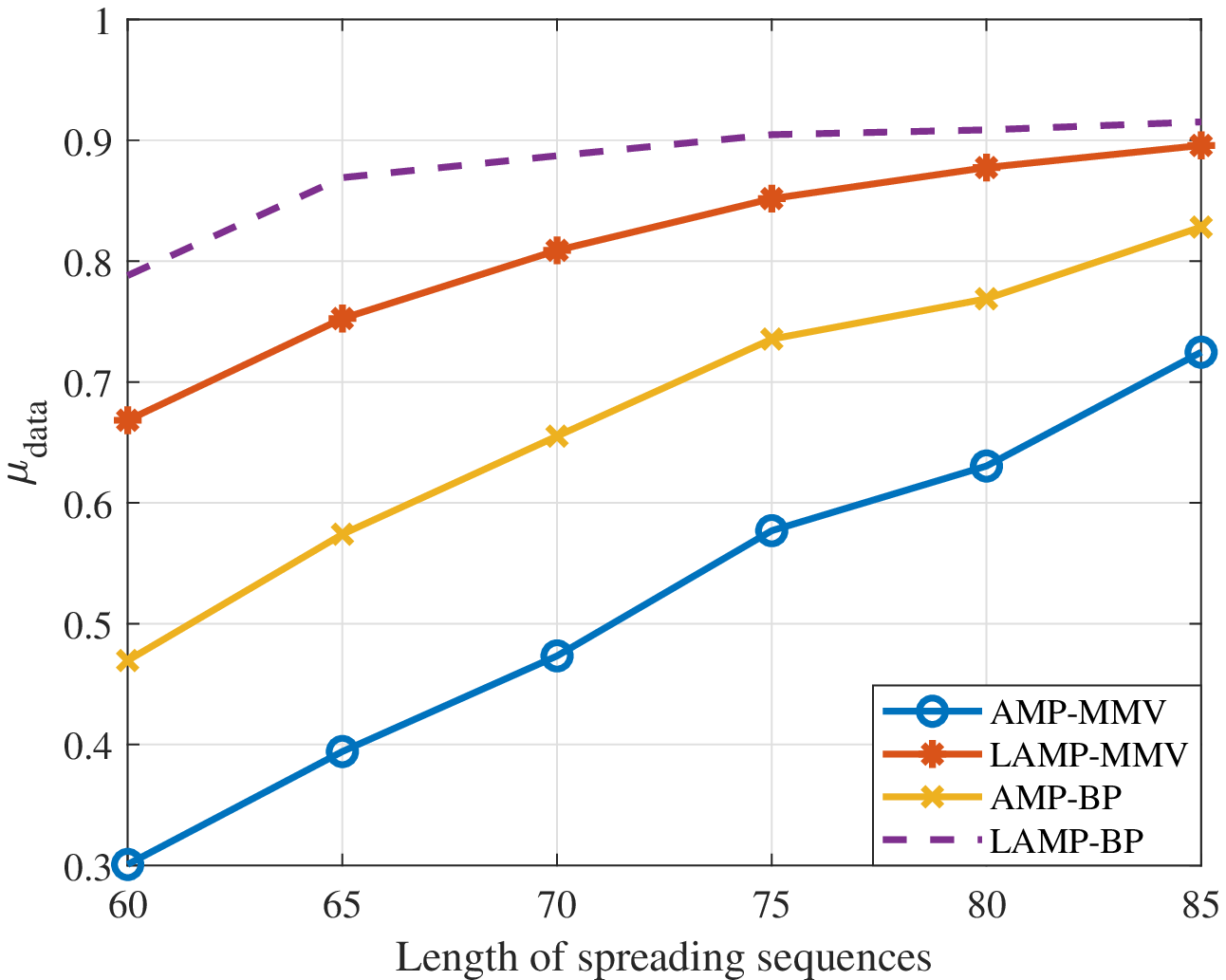}}
\caption{Pilot detection and data recovery performance under different length of spreading sequences, where $N_a=24$, $L_p=1$, $L_d=3$, $T_g=3$, $R=1$, SNR$=30dB$.}
\label{fig:diffls} 
\end{figure}
In Fig.~\ref{fig:diffls}, we compare the performance of pilot detection and data recovery under different lengths of spreading sequences. Generally, with longer spreading sequences, we need to consume more resources for pilot and data transmission, and thus obtain higher data recovery rate. The proposed LAMP-BP has the best pilot detection and data recovery performance. Therefore, under the requirement of the same data recovery rate, the proposed algorithm consumes the least resources. Besides, the proposed method shows more performance gain compared to the traditional CS algorithm. Similar mechanism can also be extended to other CS algorithms.

\begin{figure}
\centering
\subfigure[]{
\includegraphics[width=0.48\textwidth]{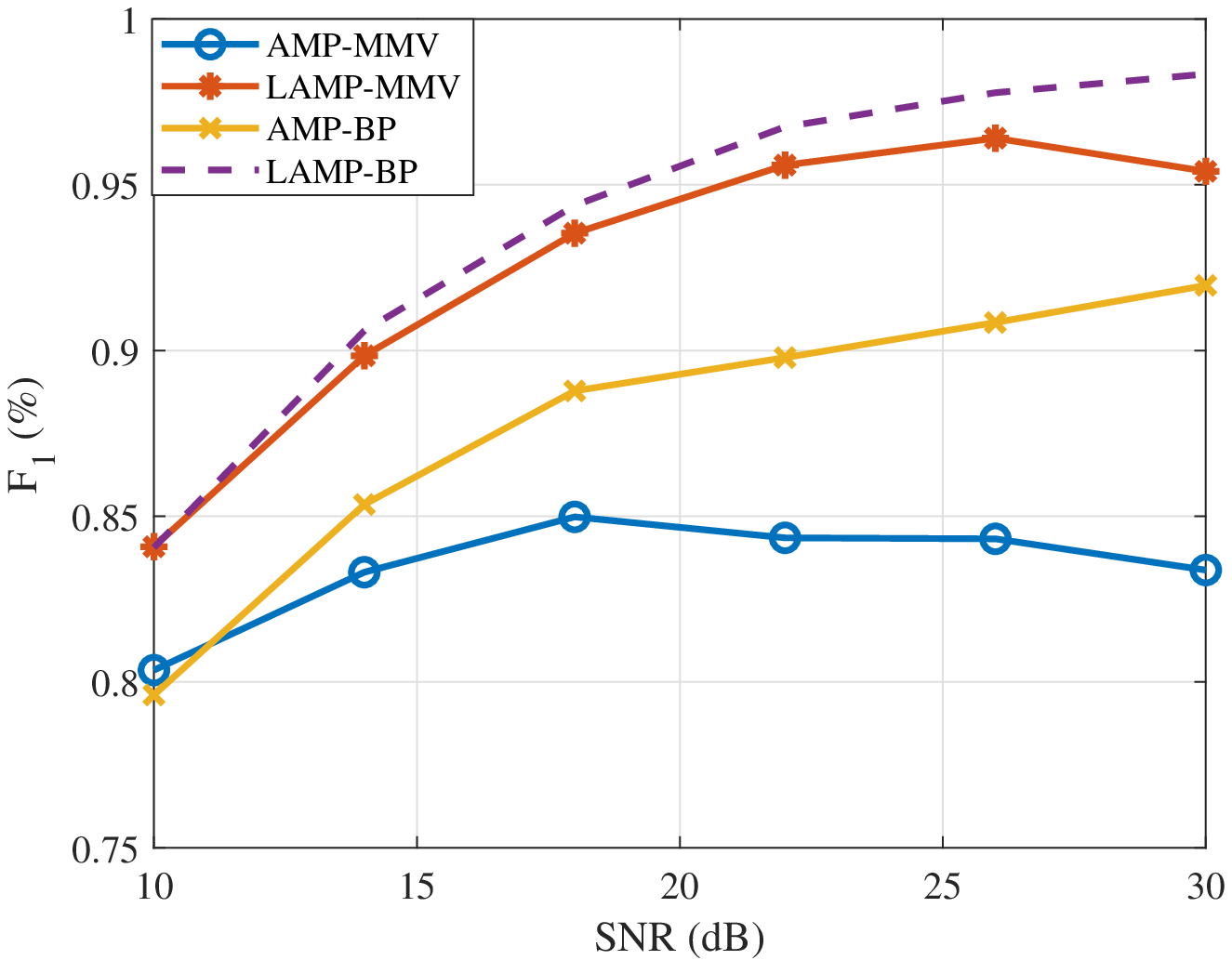}}
\vspace{.1in}
\subfigure[]{
\includegraphics[width=0.48\textwidth]{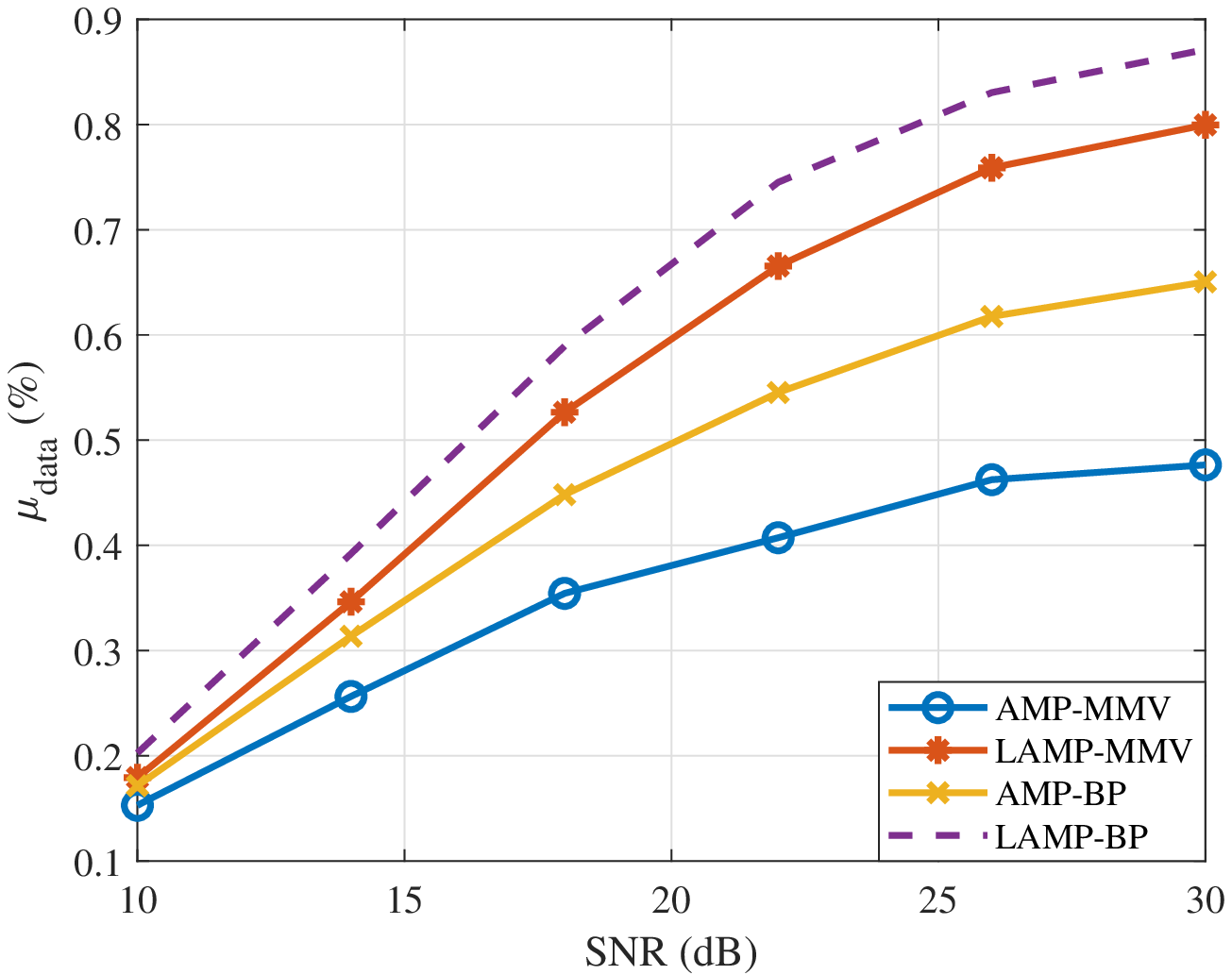}}
\caption{Pilot detection and data recovery performance under different SNRs, where $L_s=70$, $N_a=24$, $L_p=1$, $L_d=3$, $T_g=3$, $R=1$.}
\label{fig:diffsnr} 
\end{figure}
\begin{figure}
\centering
\subfigure[]{
\includegraphics[width=0.48\textwidth]{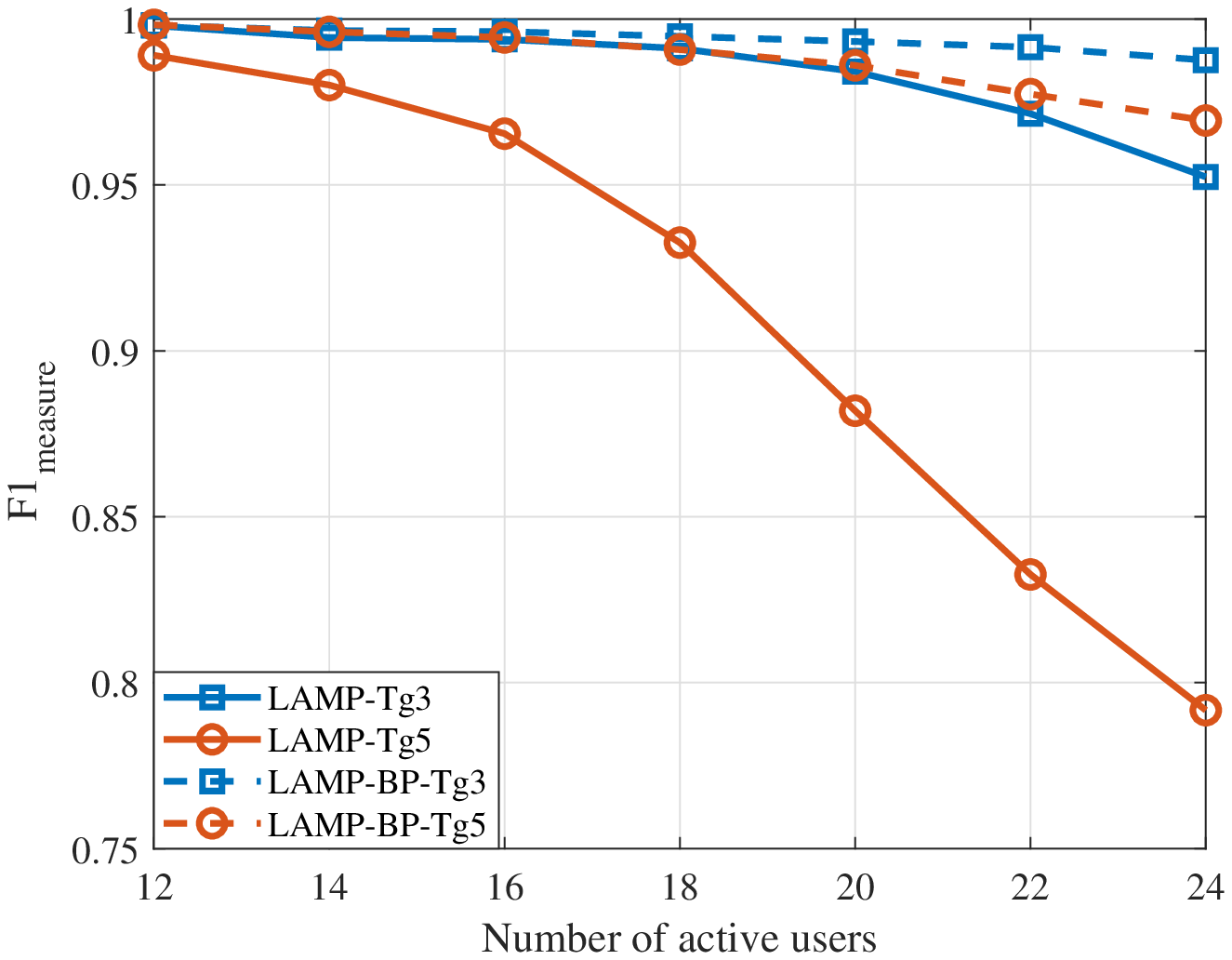}}
\vspace{.1in}
\subfigure[]{
\includegraphics[width=0.48\textwidth]{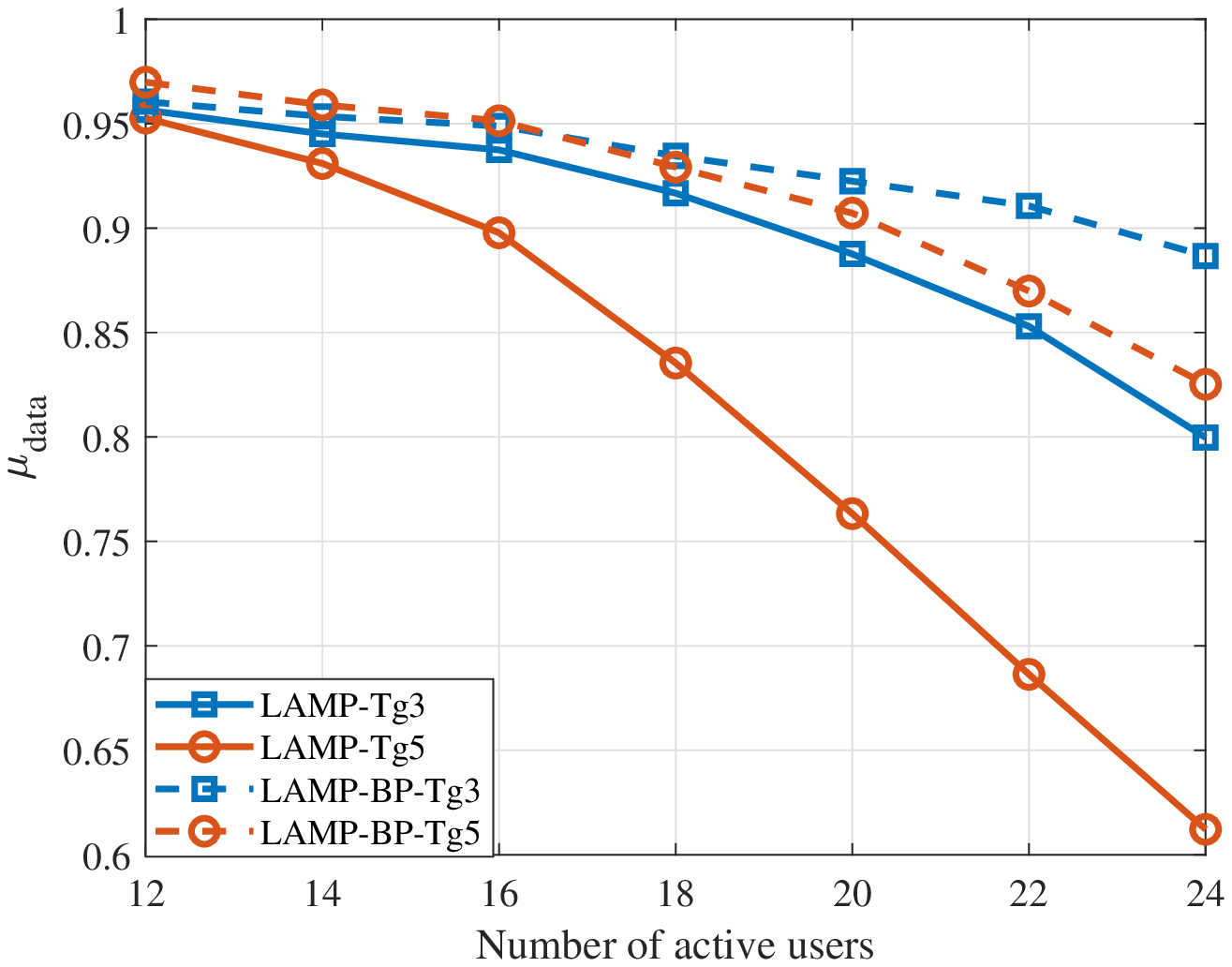}}
\caption{Pilot detection and data recovery performance of different $T_g$ under different number of active users, where $L_s=70$, $L_p=1$, $L_d=3$, $T_g=3$, $R=1$, SNR$=30dB$.}
\label{fig:difftg} 
\end{figure}

In Fig.~\ref{fig:diffsnr}, we compare the performance of pilot detection and data recovery under different SNRs. Under higher SNR, the $F_{1}$ of AMP and LAMP has a little performance degeneration. This is because the SNR is used as a threshold to determine whether a sequence is selected, and as the SNR increases, the false detection rate increases, leading to a decrease in recall, which affects the $F1_{measure}$. In Fig.~\ref{fig:difftg}, we show the detection and data recovery performance by varying the guard time. As the diversity of symbol delay increases from $3$ to $5$, the interference increases and performance decreases. In this situation, the proposed method shows a high performance gain.

\section{Conclusion}

This paper considers the asynchronous massive access of various machine-type devices which transmit different numbers of data symbols. At the receiver, the joint pilot detection, channel estimation, data recovery, and activity detection is constructed as an MMV problem with structural sparsity. The three types of sparsity come from the sporadic transmission, asynchronous access, and data length diversity. Specifically, we embed backward propagation into AMP algorithm to explore the structural sparsity, and then unfold the proposed AMP-BP into a network for more performance gain. The experimental results show that the proposed method obtains improved performances compared with AMP and LAMP. As a future work, different packet sizes can be associated with different services, coding and reliability levels, which expands the applicability of the introduced framework with data frame length diversity.

%

\ifCLASSOPTIONcaptionsoff
  \newpage
\fi

\bibliographystyle{IEEEtran}
\bibliography{refs}

\begin{thebibliography}{10}
\providecommand{\url}[1]{#1}
\csname url@samestyle\endcsname
\providecommand{\newblock}{\relax}
\providecommand{\bibinfo}[2]{#2}
\providecommand{\BIBentrySTDinterwordspacing}{\spaceskip=0pt\relax}
\providecommand{\BIBentryALTinterwordstretchfactor}{4}
\providecommand{\BIBentryALTinterwordspacing}{\spaceskip=\fontdimen2\font plus
\BIBentryALTinterwordstretchfactor\fontdimen3\font minus
  \fontdimen4\font\relax}
\providecommand{\BIBforeignlanguage}[2]{{%
\expandafter\ifx\csname l@#1\endcsname\relax
\typeout{** WARNING: IEEEtran.bst: No hyphenation pattern has been}%
\typeout{** loaded for the language `#1'. Using the pattern for}%
\typeout{** the default language instead.}%
\else
\language=\csname l@#1\endcsname
\fi
#2}}
\providecommand{\BIBdecl}{\relax}
\BIBdecl

\bibitem{8766143}
Z.~{Zhang}, Y.~{Xiao}, Z.~{Ma}, M.~{Xiao}, Z.~{Ding}, X.~{Lei}, G.~K.
  {Karagiannidis}, and P.~{Fan}, ``{6G} wireless networks: Vision,
  requirements, architecture, and key technologies,'' \emph{IEEE Vehicular
  Technology Magazine}, vol.~14, no.~3, pp. 28--41, 2019.

\bibitem{9464917}
S.~{Shahzadi}, M.~{Iqbal}, and N.~R. {Chaudhry}, ``{6G} vision: Toward future
  collaborative cognitive communication (3c) systems,'' \emph{IEEE
  Communications Standards Magazine}, vol.~5, no.~2, pp. 60--67, 2021.

\bibitem{9320554}
A.~{Azari}, \u{C}. {Stefanovi\'{c}}, P.~{Popovski}, and C.~{Cavdar},
  ``Energy-efficient and reliable iot access without radio resource
  reservation,'' \emph{IEEE Transactions on Green Communications and
  Networking}, vol.~5, no.~2, pp. 908--920, 2021.

\bibitem{9537931}
J.~Choi, J.~Ding, N.-P. Le, and Z.~Ding, ``Grant-free random access in
  machine-type communication: Approaches and challenges,'' \emph{IEEE Wireless
  Communications}, vol.~29, no.~1, pp. 151--158, 2022.

\bibitem{9097306}
M.~B. {Shahab}, R.~{Abbas}, M.~{Shirvanimoghaddam}, and S.~J. {Johnson},
  ``Grant-free non-orthogonal multiple access for {IoT}: A survey,'' \emph{IEEE
  Communications Surveys Tutorials}, vol.~22, no.~3, pp. 1805--1838, 2020.

\bibitem{9031550}
A.~C. Cirik, N.~M. Balasubramanya, L.~Lampe, G.~Vos, and S.~Bennett, ``Toward
  the standardization of grant-free operation and the associated noma
  strategies in 3gpp,'' \emph{IEEE Communications Standards Magazine}, vol.~3,
  no.~4, pp. 60--66, 2019.

\bibitem{8482464}
Y.~Du, C.~Cheng, B.~Dong, Z.~Chen, X.~Wang, J.~Fang, and S.~Li,
  ``Block-sparsity-based multiuser detection for uplink grant-free noma,''
  \emph{IEEE Transactions on Wireless Communications}, vol.~17, no.~12, pp.
  7894--7909, 2018.

\bibitem{9140386}
S.~Jiang, X.~Yuan, X.~Wang, C.~Xu, and W.~Yu, ``Joint user identification,
  channel estimation, and signal detection for grant-free noma,'' \emph{IEEE
  Transactions on Wireless Communications}, vol.~19, no.~10, pp. 6960--6976,
  2020.

\bibitem{8454392}
L.~Liu, E.~G. Larsson, W.~Yu, P.~Popovski, C.~Stefanovic, and E.~de~Carvalho,
  ``Sparse signal processing for grant-free massive connectivity: A future
  paradigm for random access protocols in the internet of things,'' \emph{IEEE
  Signal Processing Magazine}, vol.~35, no.~5, pp. 88--99, 2018.

\bibitem{1614066}
D.~Donoho, ``Compressed sensing,'' \emph{IEEE Transactions on Information
  Theory}, vol.~52, no.~4, pp. 1289--1306, 2006.

\bibitem{4385788}
J.~A. Tropp and A.~C. Gilbert, ``Signal recovery from random measurements via
  orthogonal matching pursuit,'' \emph{IEEE Transactions on Information
  Theory}, vol.~53, no.~12, pp. 4655--4666, 2007.

\bibitem{9903376}
Y.~Bai, W.~Chen, F.~Sun, B.~Ai, and P.~Popovski, ``Data-driven compressed
  sensing for massive wireless access,'' \emph{IEEE Communications Magazine},
  vol.~60, no.~11, pp. 28--34, 2022.

\bibitem{9605579}
Y.~Bai, W.~Chen, B.~Ai, Z.~Zhong, and I.~J. Wassell, ``Prior information aided
  deep learning method for grant-free noma in mmtc,'' \emph{IEEE Journal on
  Selected Areas in Communications}, vol.~40, no.~1, pp. 112--126, 2022.

\bibitem{9685696}
Y.~Bai, W.~Chen, Y.~Ma, N.~Wang, and B.~Ai, ``Dual-net for joint channel
  estimation and data recovery in grant-free massive access,'' in \emph{2021
  IEEE Global Communications Conference (GLOBECOM)}, 2021, pp. 1--6.

\bibitem{9252937}
W.~Chen, B.~Zhang, S.~Jin, B.~Ai, and Z.~Zhong, ``Solving sparse linear inverse
  problems in communication systems: A deep learning approach with adaptive
  depth,'' \emph{IEEE Journal on Selected Areas in Communications}, vol.~39,
  no.~1, pp. 4--17, 2021.

\bibitem{7934066}
M.~Borgerding, P.~Schniter, and S.~Rangan, ``Amp-inspired deep networks for
  sparse linear inverse problems,'' \emph{IEEE Transactions on Signal
  Processing}, vol.~65, no.~16, pp. 4293--4308, 2017.

\bibitem{8323218}
L.~Liu and W.~Yu, ``Massive connectivity with massive mimo¡ªpart i: Device
  activity detection and channel estimation,'' \emph{IEEE Transactions on
  Signal Processing}, vol.~66, no.~11, pp. 2933--2946, 2018.

\bibitem{9566698}
W.~Chen, H.~Xiao, L.~Sun, and B.~Ai, ``Joint activity detection and channel
  estimation in massive mimo systems with angular domain enhancement,''
  \emph{IEEE Transactions on Wireless Communications}, vol.~21, no.~5, pp.
  2999--3011, 2022.

\bibitem{9839006}
Y.~Bai, W.~Chen, Y.~Bai, and B.~Ai, ``Dictionary learning based channel
  estimation and activity detection for mmtc with massive mimo,'' in \emph{ICC
  2022 - IEEE International Conference on Communications}, 2022, pp.
  2272--2277.

\bibitem{9364871}
J.-C. Jiang and H.-M. Wang, ``Massive random access with sporadic short
  packets: Joint active user detection and channel estimation via sequential
  message passing,'' \emph{IEEE Transactions on Wireless Communications},
  vol.~20, no.~7, pp. 4541--4555, 2021.

\bibitem{9268113}
H.~Xiao, W.~Chen, J.~Fang, B.~Ai, and I.~J. Wassell, ``A grant-free method for
  massive machine-type communication with backward activity level estimation,''
  \emph{IEEE Transactions on Signal Processing}, vol.~68, pp. 6665--6680, 2020.

\bibitem{9413870}
L.~Liu and Y.-F. Liu, ``An efficient algorithm for device detection and channel
  estimation in asynchronous iot systems,'' in \emph{ICASSP 2021 - 2021 IEEE
  International Conference on Acoustics, Speech and Signal Processing
  (ICASSP)}, 2021, pp. 4815--4819.

\bibitem{8716690}
T.~Ding, X.~Yuan, and S.~C. Liew, ``Sparsity learning-based multiuser detection
  in grant-free massive-device multiple access,'' \emph{IEEE Transactions on
  Wireless Communications}, vol.~18, no.~7, pp. 3569--3582, 2019.

\bibitem{9814672}
A.~B. Baral, W.~Namgoong, and M.~Torlak, ``Joint sparse support recovery for
  asynchronous multicarrier modulation signals in cognitive radio networks,''
  in \emph{2022 IEEE International Conference on Communications Workshops (ICC
  Workshops)}, 2022, pp. 699--704.

\bibitem{9390399}
W.~Zhu, M.~Tao, X.~Yuan, and Y.~Guan, ``Deep-learned approximate message
  passing for asynchronous massive connectivity,'' \emph{IEEE Transactions on
  Wireless Communications}, vol.~20, no.~8, pp. 5434--5448, 2021.

\bibitem{9267798}
Y.~Bai, W.~Chen, B.~Ai, and Z.~Zhong, ``Contention-based nonorthogonal massive
  access with massive mimo,'' \emph{China Communications}, vol.~17, no.~11, pp.
  79--90, 2020.

\bibitem{6692494}
H.~F. Schepker, C.~Bockelmann, and A.~Dekorsy, ``Coping with cdma
  asynchronicity in compressive sensing multi-user detection,'' in \emph{2013
  IEEE 77th Vehicular Technology Conference (VTC Spring)}, 2013, pp. 1--5.

\bibitem{1453780}
S.~Cotter, B.~Rao, K.~Engan, and K.~Kreutz-Delgado, ``Sparse solutions to
  linear inverse problems with multiple measurement vectors,'' \emph{IEEE
  Transactions on Signal Processing}, vol.~53, no.~7, pp. 2477--2488, 2005.

\end{thebibliography}

\end{document}